\documentclass[12pt]{iopart}

\usepackage{epsf}
\usepackage{iopams}
\usepackage{amsfonts}
\usepackage{amssymb}
\usepackage{bm}
\usepackage{bbm}
\usepackage{nicefrac}
\usepackage{diagbox}

\usepackage{tikz}

\usepackage{soul}

\usepackage{cite}

\usepackage[breaklinks=true,colorlinks=true,linkcolor=blue,%
urlcolor=blue,citecolor=blue]{hyperref}
\usepackage{graphicx}
\usepackage{epsfig}
\usepackage{latexsym}
\usepackage{dcolumn}
\usepackage{pifont}
\usepackage{xcolor}
\usepackage{siunitx}

\usepackage{cite}

\usepackage{epsfig}
\usepackage{latexsym}

\usepackage{dcolumn}
\usepackage{pifont} 

\newcommand{\dd}{\mathrm{d}}
\newcommand{\ee}{\mathrm{e}}
\newcommand{\ii}{\mathrm{i}}

\makeatletter

\newcommand{\Rmnum}[1]{\expandafter\@slowromancap\romannumeral #1@}

\definecolor{garrosgreen}{rgb}{0.1, 0.4, 0.1}
\definecolor{dartmouthgreen}{rgb}{0.05, 0.5, 0.06}
\definecolor{duelferred}{rgb}{0.7, 0.2, 0.1}
\definecolor{cambridgeblue}{rgb}{0.1, 0.3, 1.0}
\definecolor{oxfordblue}{rgb}{0.05, 0.2, 0.7}

\newcolumntype{.}{D{x}{}{-1}}

\begin{document}

\title[{Pressure Shifts in High--Precision Hydrogen Spectroscopy: II.}]
{Pressure Shifts in High--Precision Hydrogen Spectroscopy: II.
Impact Approximation and Monte--Carlo Simulations}

\newcommand{\addrROLLA}{Department of Physics,
Missouri University of Science and Technology,
Rolla, Missouri 65409, USA}

\newcommand{\addrCHEM}{Department of Chemistry,
Missouri University of Science and Technology,
Rolla, Missouri 65409, USA}

\newcommand{\addrHDphiltheo}{Institut f\"ur Theoretische Physik,
Universit\"{a}t Heidelberg,
Philosophenweg 16, 69120 Heidelberg, Germany}

\newcommand{\addrLEBEDEV}{P  N  Lebedev Physics
Institute, Leninsky prosp.~53, Moscow, 119991 Russia}

\newcommand{\addrMUC}{Max--Planck--Institut f\"ur
Quantenoptik, Hans--Kopfermann-Stra\ss{}e~1,
85748 Garching, Germany}

\newcommand{\addrRQC}{Russian Quantum Center,
Business--Center ``Ural'', 100A Novaya street,
Skolkovo, Moscow, 143025 Russia}

\newcommand{\addrBUDKER}{Budker Institute of Nuclear Physics,
630090 Novosibirsk, Russia}

\author{A Matveev$^{1,2}$, N Kolachevsky$^{1,3}$,  
C M Adhikari$^4$\\
and U D Jentschura$^4$}
\address{$^1$\addrLEBEDEV}
\address{$^2$\addrMUC}
\address{$^3$\addrRQC}
\address{$^4$\addrROLLA}

\begin{abstract}
We investigate collisional shifts of 
spectral lines involving excited hydrogenic states,
where van der Waals coefficients have 
recently been shown to have large numerical values
when expressed in atomic units.
Particular emphasis is laid on the recent hydrogen $2S$--$4P$ experiment
(and an ongoing $2S$--$6P$ experiment) in Garching,
but numerical input data are provided for other transitions
(e.g., involving $S$ states), as well.
We show that the frequency shifts can be described,
to sufficient accuracy, in the impact approximation.
The pressure related effects were separated into two parts, 
{\em (i)} related to collisions of atoms inside of the beam,
and {\em (ii)} related to collisions of the atoms 
in the atomic beam with the residual background gas.
The latter contains both 
atomic as well as molecular hydrogen.
The dominant effect of intra-beam
collisions is evaluated by a Monte-Carlo simulation,
taking the geometry of the experimental apparatus into account. 
While, in the 
Garching experiment, the collisional shift is on the order of 10\,Hz,
and thus negligible, it can decisively depend on the experimental conditions.
We present input data which can be used 
in order to describe the effect for other transitions
of current and planned experimental interest.
\end{abstract}

\pacs{31.30.jh, 31.30.J-, 31.30.jf}

\vspace{2pc}
\noindent{\it Keywords}: Collisional Shift; Collisional Broadening;
van der Waals Interaction; Impact Approximation, Monte--Carlo Approach



%
%
\section{Introduction}

High-precision spectroscopy experiments 
on atomic hydrogen~\cite{BeEtAl2017,Fl2017,%
HoHo2010,YoEtAl2016,MaEtAl2013prl}
are critically important sources of data for the 
least-square adjustment of
fundamental constants~\cite{MoNeTa2016}. The discrepancy in 
the interpretation of the
results of related experiments,
notably, in extracting the proton charge 
radius from ordinary hydrogen versus 
muonic hydrogen (known as the proton size puzzle,
see Refs.~\cite{PoEtAl2010,PoEtAl2016}), raises questions concerning 
conceivable systematic effects which can be
overlooked in experiments. Among these,
pressure-related effects (collisional shifts) need to be studied in more detail. 

The absence of efficient laser-cooling techniques
for atomic hydrogen makes it very difficult to
devise collision-free methods of spectroscopy, e.g.,
those based on optical lattices. A
standard method, which may be used for 
an immediate experimental evaluation of collisional
shifts, is based on the variation of the pressure
(extrapolating the spectroscopic
results to vanishing particle density). However,
the extrapolation procedure is 
also connected with an uncertainty, which 
affects the resulting
uncertainty of the experiment. 
In general, if the magnitude of 
the pressure-related shifts in the
experiment is smaller than or comparable to its overall 
uncertainty (due to other effects), then it is difficult
to use an extrapolation procedure 
effectively. Under these conditions,
theoretical estimates
of the collisional shift become indispensable.
The simplicity of the hydrogen atom helps
in this regard.

From a historical perspective, it is interesting to 
note that the study of collisional shifts and broadening mechanisms was started 
more than a century ago~\cite{Mi1895, Lo1906}. 
However, the application of the
developed methods to precision spectroscopy requires some efforts. The
spectroscopy of atomic hydrogen takes place in 
atomic beams, where the
distribution of relative velocities between atoms cannot be described by a 
simple function, like the Maxwell distribution. The geometry of the atomic beam and the
velocity-selectivity of the data acquisition may also affect the evaluation. 
 Because of the $1/R^6$ dependence of the 
van der Waals interaction of atoms, the effective range of the 
interatomic interaction is limited to a few hundred Bohr
radii, and the collisions happen very fast when 
measured in terms of typical lifetimes of 
excited atomic states. This observation justifies, as
we shall discuss in detail in the following,
the so-called impact approximation~\cite{SoVaYu1981}
which describes the effect of the collisions 
as sequences of ``sudden'' phase shifts,
which in turn depend on the impact parameter and on the 
velocity of the atoms. 
Hence, details of the experimental apparatus need to be 
considered in the theoretical calculation.

The goal of this paper is to describe in detail,
a procedure for the estimation of both
frequency shifts as well as line broadening in 
precision atomic-beam measurements of transitions
in hydrogen and other simple atomic systems. 
Particular focus will be laid 
on the collisional shift 
in the recently completed $2S$--$4P$ experiment,
on a beam of cold atomic
hydrogen in Garching \cite{BeEtAl2017}. 
However, we emphasize that the results of this work can also be 
used for future experiments on spectroscopy of other $2S$--$nP$ transitions on
the same apparatus (e.g., for $n=6$),
and with minor modifications, 
for other transitions which will be the focus of 
attention in the future.
Recent progress in the determination of interaction
potentials between neutral hydrogen atoms
and higher excited hydrogen atoms 
and molecules~\cite{AdEtAl2017vdWi,
JeEtAl2017vdWii,JeAdDaMaKo2018jpb1} opens the 
possibility for an improved calculation of the 
collisional cross-sections and the corresponding shift and 
line broadening.

In beam spectroscopy experiments,
it is convenient to separately consider 
the collisions of the atoms inside the beam with each other 
(intra-beam collisions) and collisions of the atoms 
with the background gas. 
The beam-background shift is related to the pressure 
of the background gas, while the 
intra-beam shift is related to the flux of gas. 
The estimation of the intra-beam shift can in principle be done 
analytically, and supplemented by 
a Monte-Carlo approach. The beam-background collisional shift 
can be estimated analytically,
using as input the residual pressure of background gas 
in the vacuum chamber, which in our case 
is better than $10^{-8}$\,mbar. 

We organize this paper as follows: In Sec.~\ref{sec2}, we 
present a brief discussion of the basic physical 
ideas, and of the impact approximation used for our analysis. In
Sec.~\ref{sec3}, we derive the cross-sections for 
the collisional shift in H--H
collisions, using recently obtained results for the long-range 
van der Waals interaction coefficients. The
calculation of the collisional shift for the already mentioned
$2S-4P$ experiment and estimation of this effect for a possible upcoming
$2S-6P$ transition measurement 
are completed in Sec.~\ref{sec4}. 
Conclusions are drawn in Sec.~\ref{sec5}.
SI mksA units are used throughout the paper,
in order to enhance the readability and reproducibility 
of the obtained results.

%
%
\section{Impact approximation}
\label{sec2}

A standard method to find the pressure shift in a rarified gas 
is based on the 
so-called impact approximation (see Chap.~36 of Ref.~\cite{So1972}). 
The main assumption of this approximation is that we can neglect 
the interaction of the spectator atom with the 
perturbing species except for a very short period of time,
when perturbing atoms approach the spectator closely.
For this approximation to be valid, the 
collision has to happen on time scales short
compared to the natural lifetime of the 
excited atomic state.
In the framework of the impact approximation, 
we thus neglect the duration of the collision $\tau_{\rm col}$
and consider the process as instantaneous,
assuming that $\tau_{\rm col}\ll\Gamma^{-1}$, where 
$\Gamma$ is the decay constant (imaginary part of the 
excited-state energy) of the atomic levels.
For example, the collision time in the Garching $2S$--$4P$ 
experiment can be estimated
in terms of the so-called Weisskopf radius, which is 
a critical value of an impact parameter where the 
phase change during a collision reaches 
the value of unity. (In general, a larger Weisskopf radius 
implies a stronger interatomic interaction.)
The Weisskopf radius for the 
Garching $2S$--$4P$ experiment is
less or on the order of $100\,a_0$, and the collision velocity is in 
the order of $300$\,m/s.
Thus, $\tau_{\rm col} \sim 10^{-10}$ s, while the
natural lifetime of the $4P$ state is about 
$1.24 \times 10^{-8}$\,s~\cite{BeSa1957}, 
which justifies the use of the impact approximation.

Let us consider a two-level atom with an initial state 
$|g\rangle$ and an excited state $|e\rangle$,
and an energy difference 
between those levels equal to $\hbar \omega_0$. The free evolution of the 
off-diagonal matrix element between those states can be written as 
$\rho_{ge}(t)=\rho_{ge}(0) \exp(-\ii \, \omega_0 \, t)$. The collisions with other atoms affect
the phase of the oscillations, causing a drift of the phase, which we can associate 
with a shift and a broadening of the line via a Fourier transformation.
So, we can write the oscillating term of the off-diagonal matrix element as:
\begin{eqnarray}
f(t)= \exp[ -\mathrm{i}\, \omega_0 t-\mathrm{i} \, \psi(t) ] \,,
\end{eqnarray}
where $\psi$ is a random function, describing the total phase
acquired during a collision. 
Within our approximations, the collision happens instantaneously, 
and we can write
\begin{eqnarray}
\psi(t) = \sum_i \phi_i \, \Theta(t-t_i),
\end{eqnarray}
where $\Theta$ is the Heaviside step function, 
$\phi_i$ is a phase shift gained in 
$i$th collision, while $t_i$ is the time of the $i$th event.
The autocorrelation function of this oscillating process~is
\begin{eqnarray}
\label{autocorr}
A(\tau)=\langle f(t)f^*(t-\tau)  \rangle 
= \ee^{-\ii \omega_0 \tau} \langle 
\exp[ \, -\ii \, \left\{ \psi(t)-\psi(t-\tau) \right\} \, ] \rangle,
\end{eqnarray}
where $\langle...\rangle$ represents the averaging on time axis. 
We will assume that all collisions happen independently of each other
and the distributions do not posses memory.
This corresponds to a Markov 
process, which can be described by Poisson statistics.
The probability of a collision with phase shift 
$\phi$ can be described by introducing a 
density function $a(\phi)$,
whose physical meaning is that 
\begin{eqnarray}
\label{defa}
\dd p = a(\phi) \, \dd t \, \dd \phi
\end{eqnarray}
is the probability 
of collision with a phase shift between $\phi$ and 
$\phi+\dd \phi$ during the time interval $\dd t$.
According to the Poisson distribution,
the number of the collisions $k$ with a phase shift in 
the interval  $(\phi,\phi+\dd \phi)$ 
during the time $\tau$ is distributed as
\begin{eqnarray}
p(k)=\frac{\lambda^{k}}{k!}e^{-\lambda},
\qquad
\qquad
\lambda = a(\phi) \, \tau \, \dd \phi \,.
\end{eqnarray}
The average value 
$\langle \exp(-\mathrm{i} k \phi) \rangle$ can be computed easily,
\begin{eqnarray}
\label{pre}
\langle \exp(-\mathrm{i} k \phi) \rangle
= \sum_{k=0}^\infty p(k) e^{-\mathrm{i} k \phi}
= \sum_{k=0}^{\infty } \frac{\lambda^k e^{-\lambda-\mathrm{i} k \phi }}{k!}
= \exp\left[ -\lambda(1 - \ee^{-\ii \phi}) \right] \,. 
\end{eqnarray}
In order to continue, we discretize the space of $\phi$ to the finite set 
of values $\phi_1, \phi_2,\dots,\phi_q$, paying attention that in the end,
we need to study the behavior of the equation in the limit of 
$q\rightarrow\infty$. It is 
easy to show that when we consider collisions with different values of 
$\phi$, this formula can be generalized to the form
\begin{eqnarray}\label{NetAvg}
\langle \ee^{-\mathrm{i}(\psi(t)-\psi(t-\tau))} \rangle 
= \left< \exp\left(-\mathrm{i}\sum_{j=1}^q k_j\phi_j\right) \right>
= \left<\prod_{j=1}^{q}\exp \left(-\mathrm{i}k_j\phi_j\right)\right> \nonumber\\
= \sum_{k_1=0}^\infty p(k_1)\sum_{k_2=0}^\infty p(k_2)\cdots
\sum_{k_j=0}^\infty p(k_j)\cdots 
\sum_{k_q=0}^\infty p(k_q)\,\prod_{j=1}^{q}\exp \left(-\mathrm{i}k_j\phi_j\right)\nonumber\\
= \prod_{j=1}^{q}\sum_{k_j=0}^\infty p(k_j)\exp \left(-\mathrm{i}k_j\phi_j\right)
= \prod_{j=1}^{q}\exp\left(-\lambda_j\right)\sum_{k_j=0}^\infty 
\frac{\lambda_j^{k_j}}{k_j!}\exp \left(-\mathrm{i}k_j\phi_j\right)\,.
\end{eqnarray}
With the help of Eq.~(\ref{pre}),
we may express this as
\begin{eqnarray}\label{AvgExprn}
\langle \ee^{-\ii (\psi(t)-\psi(t-\tau))}\rangle 
&=&
\prod_{j=1}^q \left< \exp( -\ii k_j \phi_j ) \right> 
= \prod_{j=1}^{q}
\exp\left[ -\lambda_j \,
\left(1 - \exp\left(-\ii \phi_j\right)\right) \right]
\nonumber\\
&=&\exp\left(-\sum_{j=1}^q \lambda_j\left(1-
 \exp\left(-\mathrm{i}\phi_j\right)\right)\right)\,.
\end{eqnarray}
In the limit of $q$ approaching  infinity, one can express the  
sum on the right-hand side of Eq.~(\ref{AvgExprn}),
in integral form, as
\begin{eqnarray}\label{Avg:qtoInfty}
\langle  e^{-\mathrm{i}(\psi(t)-\psi(t-\tau))} \rangle 
= \exp\left(-\tau \int_{-\infty}^\infty a(\phi) \, [1-\ee^{-\ii \, \phi}]\, 
\mathrm{d}\phi \right)\nonumber\\
=\exp\left(-\tau \int_{-\infty}^\infty 
a(\phi)[1-\cos\phi]\,\dd \phi \right) \;
\exp\left( -\ii \, \tau \int_{-\infty}^\infty 
a(\phi) \, \sin\phi\, \dd \phi \right)\,,
\end{eqnarray}
where we make use of $\lambda = a(\phi) \, \tau \, \dd \phi$.
The phase factor has to be added to the 
phase $\exp(-\ii \omega_0 t)$ from Eq.~(\ref{autocorr}).
According to the Wiener-Khinchin theorem, the power spectrum of the process 
$f(t)$ can be obtained by Fourier transform of the autocorrelation function. 
This calculation gives a Lorentz function
\begin{eqnarray}
\tilde{f}(\omega) \sim \frac{(\gamma_c/2)^2}{(\gamma_c/2)^2
+(\omega-\omega_0-\omega_c)^2},
\end{eqnarray}
where $\omega_c$ and $\gamma_c$ are 
obtained as the collisional shift and broadening:
\begin{eqnarray}
\label{omegac_res}
\omega_c &=& \int_{-\infty}^{+\infty} 
a(\phi) \, \sin(\phi) \, \mathrm{d}\phi\,,
\\[0.1133ex]
\label{gammac_res}
\gamma_c &=& \int_{-\infty}^{+\infty} 
a(\phi) \, [1-\cos(\phi)] \, \mathrm{d}\phi\,.
\end{eqnarray}
The physical dimension of these equations can easily be checked
upon observing that $a(\phi)$, according to Eq.~(\ref{defa}),
carries a physical dimension of inverse time.

%
%
\section{Calculation of collisional shifts and broadenings}
\label{sec3}

%
%
\subsection{Peculiarities of van der Waals coefficients for $S$ states}
\label{sec31}

Even if the main subject of the current paper is the pressure
shift in $1S$--$nP$ transition, we here recall a few
interesting aspects of calculations related to 
van der Waals interactions of atomic hydrogen atoms in $S$ states.
We consider a two-atom system in which both atoms $A$ and $B$  
are in $|kS\rangle$ and $|nS\rangle$, where $k, n \in \mathbb{N}$. 
The energetic degeneracy of states $|kS\rangle_A |nS\rangle_B$
and $|nS\rangle_A |kS\rangle_B$ indicates that the $kS$--$nS$  
exchange interaction results in entanglement of the states, 
whose basis states are 
\begin{eqnarray}
|\Psi_1\rangle=|kS\rangle_A |nS\rangle_B
\qquad \mbox{and} \qquad
|\Psi_2\rangle=|nS\rangle_A |kS\rangle_B\,.
\end{eqnarray}
The eigenvalue equation of the system is given by 
\begin{eqnarray}
\left(H_0+H_{\rm vdW}\right) |\Psi\rangle=E |\Psi\rangle\,,
\end{eqnarray}
where the unperturbed 
Hamiltonian $H_0$ is the sum of the Schr{\"o}dinger Hamiltonians 
of the atoms,
\begin{eqnarray}
H_0=\sum_{i=A,B}\left(\frac{\vec{p}_i^{\,2}}{2m} - 
\frac{e^2}{4\pi\epsilon_0}\frac{1} {|\vec{r}_i|}\right)\,.
\end{eqnarray}
Here, $m$ is the mass of the electron, while 
$\vec{p}_i$ and $\vec{r}_i$ are the kinetic momenta
of the electron for an  atom $i$ and the position of the electron 
relative to the  nucleus  of the atom,  respectively. 
The van der Waals Hamiltonian,
\begin{eqnarray}
\label{vdw}
H_{\rm vdW} = \frac{e^2}{4 \pi \epsilon_0} \, 
\frac{x_A \, x_B + y_A \, y_B - 2 \, z_A \, z_B}{R^3} \,,
\end{eqnarray}
is the perturbation to the system.  
Here $x_i$, $y_i$, and $z_i$ are the coordinates of the 
atomic electrons with respect to the atomic 
centers, while $R$ is the interatomic distance.
As given in Sec. IV 
of Ref.~\cite{AdEtAl2017vdWi}, the eigenenergies and the eigenvectors 
of the system, in the van der Waals range 
$(a_0\ll R \ll a_0/\alpha)$, can be expressed as 
\numparts
\begin{eqnarray}
\label{D6sign}
E_{\pm}= E_0 - \frac{D_6(nS;kS)\pm M_6 (nS;kS)}{R^6}\,,\\
 |\Psi\rangle= \frac{1}{\sqrt{2\,}}\left(|\Psi_1\rangle\pm |\Psi_2\rangle\right)\,,
\end{eqnarray}
\endnumparts
where $E_0$ is the unperturbed energy, 
and $D_6(nS;kS) $ and $M_6(nS;kS)$ are, respectively,
the direct and mixing van der Waals coefficients. 
They are given by Eq.~(54) of Ref.~\cite{AdEtAl2017vdWi}. 
Let us  recall them  here for convenience,
\numparts
\begin{eqnarray}
\label{sum_over_states}
D_6(nS;kS) = \frac{2\, e^4}{3\left(4 \pi \epsilon_0\right)^2}\,
\sum_{pq} \mkern -23.3mu \int \frac{\left| \langle kS|\vec{r} \,|p\rangle\right|^2
\left| \langle nS|\vec{r}\, |q\rangle\right|^2}{E_p+E_q-(E_{kS}+E_{nS})}\;,\\
M_6(nS;kS) = \frac{2\, e^4}{3\left(4 \pi \epsilon_0\right)^2}
\sum_{pq} \mkern -23.3mu \int \frac{ \langle kS|\vec{r}\, |p\rangle\cdot
\langle p | \vec{r} | nS\rangle
\langle nS|\vec{r}\, |q\rangle\cdot \langle q | \vec{r}\, | kS\rangle}
{E_p+E_q-(E_{kS}+E_{nS})}\,.
\end{eqnarray}
\endnumparts
Here, the sum-integral sign clarifies that the 
continuum states are to be included in the sum over
virtual states, and the transition matrix 
elements are to be evaluated for the two atoms separately,
as indicated.
The symmetry-dependent quantity $D_6(nS;kS)\pm M_6 (nS;kS)$ 
is the  van der Waals coefficient of the $nS$--$kS$ system. 
More explicitly,
\begin{eqnarray}
\label{defC6}
C_6(nS;kS) = D_6(nS;kS)\pm M_6 (nS;kS)\,.
\end{eqnarray}
For example, the coefficients
$C_6(2S;1S)$ (see Ref.~\cite{AdEtAl2017vdWi})
and $C_6(3S;1S)$ (see Ref.~\cite{AdEtAl2017vdWiii})
are given as 
\numparts
\begin{eqnarray}
C_6(2S;1S) = (176.752\,266\pm 27.983\,245) \, E_h \, a_0^6,\\
C_6(3S;1S) = (917.478\,571\pm 2.998\,270) \, E_h \, a_0^6\,,
\end{eqnarray}
\endnumparts
where $E_h= \alpha^2 m c^2$ and $a_0=\hbar/(\alpha m c)$
are the Hartree energy and the Bohr radius, respectively. 
(In atomic units, both $E_h$ and $a_0$ are unity.)
Notice that, for the $2S$--$1S$ system, the $D_6$
coefficient is about six times larger than the 
$M_6$ coefficient whereas for the $3S$--$1S$ system 
the $D_6$ coefficient  is  two orders of magnitude 
larger than the $M_6$ coefficient.
The $M_6$ coefficient becomes  
smaller with the principal quantum number is being increased
as recently observed in Ref.~\cite{AdEtAl2017vdWiii}.
However, the $D_6$ coefficients increase as the 
fourth power of the principal quantum number of the 
excited reference states. As a consequence, 
the mixing $M_6$ coefficient becomes negligible in 
comparison to the direct $D_6$ coefficient for 
higher excited reference states.

%
%
\subsection{Peculiarities of van der Waals coefficients for $P$ states}
\label{sec32}

\begin{figure}[t!]
\begin{center}
\includegraphics [width=0.4\columnwidth]{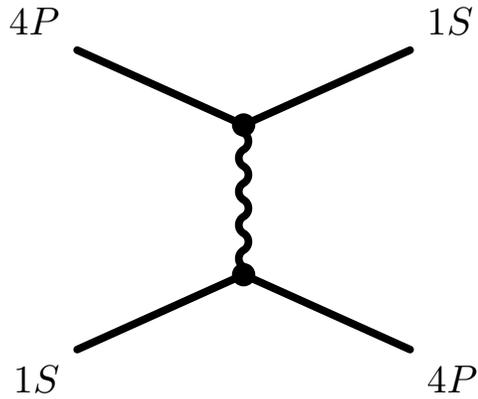}
\caption{One-photon exchange from a $(4P;1S)$ atomic hydrogen
state to a $(1S;4P)$ state, for long-range interactions between two atoms. 
\label{fig1}}
\end{center}
\end{figure}

For interactions involving higher excited $P$ states
(in atomic hydrogen), other issues arise.
Namely, in a sum-over-states representation
[see Eq.~(\ref{sum_over_states}) and~\ref{appA}],
the van der Waals $C_6$ coefficient is obtained
in terms of dipole transitions of the two atoms 
to virtual levels accessible via such transitions, 
with the sum of the transition frequencies for the virtual 
transitions of both atoms in the denominator. 
The $C_6$ coefficients in the $nP$--$1S$ and $nP$--$2S$
systems are enhanced because of the presence of 
quasi-degenerate virtual states which are accessible via such 
transitions. 
This is illustrated in Fig.~\ref{fig1} for the 
$4P$--$1S$ system: An allowed one-photon exchange 
from an initial $(1S;4P)$ state couples to the quasi-degenerate 
$(4P;1S)$ level. Therefore, hyperfine frequencies enter
the propagator denominator in second-order perturbation
theory, and the $C_6$ coefficient is drastically enhanced~\cite{JeAdDaMaKo2018jpb1}.

For example, we have according to Eq.~(69) 
of Ref.~\cite{JeAdDaMaKo2018jpb1} for the $4P$--$1S$ system, 
\numparts
\begin{eqnarray}
\label{4P12_1S}
\left< C_6(4P_{1/2}; 1S) \right> & =&
2.489 \times 10^4 \, E_h \,a_0^6, \\
\label{4P32_1S}
\left< C_6(4P_{3/2}; 1S) \right> & =&
-1.245  \times 10^4 \, E_h \,a_0^6 \,.
\end{eqnarray}
\endnumparts
Here, by $\left< C_6 \right>$, we denote
a linear average over the hyperfine manifolds 
has been performed, which is (slightly) different over 
the averaging procedure required for the calculation of the 
pressure shift (see~\ref{appB}).
Specifically, one has
$\left< | C_6 |^{2/5} \right> \neq | \left< C_6 \right> |^{2/5}$.
(Numerical experiments show that the 
two quantities differ by no more than 30\,\% for typical
atomic transitions in hydrogen.)
By contrast, for the $6P$--$1S$ system, we have according to Eq.~(27)
of Ref.~\cite{JeAd2017atoms},
\numparts
\begin{eqnarray}
\label{6P12_1S}
\left< C_6(6P_{1/2}; 1S) \right> & =&
-8.2347 \times 10^2 \, E_h \,a_0^6, \\
\label{6P32_1S}
\left< C_6(6P_{3/2}; 1S) \right> & =&
4.1174 \times 10^2 \, E_h \,a_0^6 \,,
\end{eqnarray}
\endnumparts
so the sign pattern is reversed as compared to the 
$4P$--$1S$ system.

The enhancement of $C_6$ could of course only occur if the 
two atomic or molecular species involved in the 
long-range interaction have identical or very similar
transition frequencies. Typically, this would be 
the case for identical atoms or molecules, or 
closely related ones, like different isotopes. For the reference $(4P;1S)$ system,
the virtual transition frequency to $(1S;4P)$ is almost 
zero because the two virtual transition frequencies for the two
atoms (almost) cancel. The first-order perturbation, for 
absolute degeneracy, is treated in~\ref{appA}.

%
%
\subsection{Cross sections}
\label{sec33}

In order to establish a connection between the 
gas model and the oscillator model, we can
parameterize all collisions by the relative velocity of the
colliding atoms $v$ and
the impact parameter $b$. If we write the interaction between the
spectator atom in the state $|s\rangle$ ($s=g,e$) and the perturbing 
atom at a distance $R$ as
$E_s(R)$, then the phase shift in a 
single collision can be calculated in non-recoil limit as
\begin{eqnarray}
\label{phi13}
\phi(v,b) = \frac{1}{\hbar}\int_{-\infty}^{+\infty}
\left(E_e(\sqrt{v^2t^2+b^2})-E_g(\sqrt{v^2t^2+b^2})\right)\mathrm{d}t.
\end{eqnarray}
In order to convert the formulas~(\ref{omegac_res})
and~(\ref{gammac_res}) into cross sections,
we should recall that $\dd p/\dd t = a(\phi) \, \dd \phi$,
and that 
\begin{eqnarray} 
\label{aPhi:dPhi}
\left( \frac{\dd A \, \dd p}{\dd N} \right) =
2 \pi b \, \dd b =
\left( \frac{\dd t \, \dd A}{\dd N} \right)
\left( \frac{\dd p}{\dd t} \right) =
\left( \frac{1}{\mathbbm{n} v_c} \right) \,
\left( a(\phi) \, \dd \phi \right),
\end{eqnarray}
where $\dd N$ is an infinitesimal number of
atoms, $\dd A$ is a cross-sectional
impact area, and $\mathbbm{n}$ is the number density of the atoms.
Also, $v_c$ is the velocity of collisions.
This implies that 
$a(\phi) \, \dd \phi  = \mathbbm{n} v_c \, 2 \pi b \, \dd b$,
where $v_c$ is the velocity of collisions.
Hence, we can compute the cross-sections of the 
pressure shift and broadening (in units of rad\,m$^2$):
\numparts
\begin{eqnarray}
\sigma_\omega (v) &=& 
\int_0^\infty 2\pi b \sin\left(\phi(v,b)\right) \, \dd b \,, 
\\[0.1133ex]
\sigma_\gamma (v) &=& 
\int_0^\infty 2\pi b  \, [1-\cos\left(\phi(v,b)\right)] \,
\dd b \,.
\end{eqnarray}
\endnumparts
With the help of the formulas for the cross-sections,
we estimate the shift and broadening 
of the spectral line by simple formulas
\numparts
\begin{eqnarray}
\label{omega_c}
\omega_c &=& \mathbbm{n} \, v_c \, \sigma_\omega(v_c)\,,
\\[0.1133ex]
\label{gamma_c}
\gamma_c &=& \mathbbm{n} \, v_c \, \sigma_\gamma(v_c) \,,
\end{eqnarray}
\endnumparts
where $v_c$ is a characteristic velocity of the 
collisions, which will be specified in greater detail
in the following. 
An improvement of this estimate requires a more careful 
consideration of the distribution of collisional velocities
(e.g., with the help of a Monte Carlo simulation). 

We shall also notice that in some cases the integrals in previous equations 
can be computed analytically. Particularly, if the interaction energy 
can be expressed as $E(R) = -\,C_n \, R^{-n}$, 
where $n=4,5,6,\ldots$ is a positive integer number, then
the cross-section corresponding to the pressure shift reads,
in view of Eq.~(\ref{phi13}),
\begin{eqnarray}\label{XSec:shift}
\sigma_\omega (v,n) &=& \int_0^\infty 2\pi b\sin\left(-\frac{C_n}{\hbar}
\int_{-\infty}^{+\infty} \left(b^2+t^2 v^2\right)^{-n/2}\mathrm{d}t\right)
 \mathrm{d}b\nonumber\\
&=&- \text{sgn}(C_n)
\int_0^\infty 2\pi b\sin\left(\frac{|C_n|}{\hbar\, v}\,\sqrt{\pi}\,b^{1-n}\; 
\frac{\Gamma{\left(\frac{n-1}{2}\right)}}
{ \Gamma{\left(\frac{n}{2}\right)}}  \right) \mathrm{d}b\,.
\end{eqnarray}
For a positive integer $n$, Eq.~(\ref{XSec:shift}) can be expressed as 
\begin{eqnarray}
\label{sigma_omega_sign}
\sigma_\omega (v,n) = -A_\omega(n) \;
\text{sgn}(C_n) \left(\frac{|C_n|}{\hbar v}\right)^{2/(n-1)}\,,
\end{eqnarray}
where $\text{sgn}(x)$ is the sign function, i.e.,
$\text{sgn}(x)=1$ if $x>0$ and $\text{sgn}(x)=-1$ if $x<0$. 
Since the cross-section is just a proportionality coefficient between
the flux of atoms and the experienced frequency shift, the sign of $\sigma_\omega$
deserves a remarks According to Eq.~(\ref{D6sign}),
a positive $C_6$ coefficient is associated with an 
attractive van der Waals interaction,
which in turn leads to a negative frequency shift in
Eq.~(\ref{sigma_omega_sign}).
In contrast to $\sigma_\omega (v,n)$, we shall define,
in the following, the cross section
$\sigma_\gamma (v,n)$ associated with collisional broadening,
as a manifestly positive quantity.

The coefficients 
$A_\omega(n)$ are  $n$-dependent dimensionless constants. 
For $n\ge 4$, $A_\omega(n)$ is given by 
\begin{eqnarray}
A_\omega(n)=\pi ^{n/(n-1)}\, \Gamma \left(\frac{n-3}{n-1}\right)
\left(\frac{\Gamma \left(\frac{n-1}{2}\right)}%
{\Gamma \left(\frac{n}{2}\right)}\right)^{2/(n-1)}
\sin \left(\frac{\pi }{n-1}\right)\,.
\end{eqnarray}
Similarly, the  cross-section  corresponding to the 
pressure broadening for any $n\ge 3$  reads
\begin{eqnarray}
\sigma_\gamma (v,n) &=& \int_0^\infty 2\pi b
\left[1-\cos\left(\frac{C_n}{\hbar\, v}\,\sqrt{\pi}\,b^{1-n}\; 
\frac{\Gamma{\left(\frac{n-1}{2}\right)}}
{ \Gamma{\left(\frac{n}{2}\right)}}  \right)\right] \mathrm{d}b\nonumber\\
&=& A_\gamma(n) \left(\frac{|C_n|}{\hbar v}\right)^{2/(n-1)}.
\end{eqnarray}
Both the $ A_\omega(n)$ and the $A_\gamma(n)$ coefficients 
are dimensionless. For $n>3$, we have
\begin{eqnarray}
A_\gamma(n)=\pi ^{n/(n-1)}\, 
\Gamma\left(\frac{n-3}{n-1}\right)
\left(\frac{\Gamma \left(\frac{n-1}{2}\right)}%
{\Gamma \left(\frac{n}{2}\right)}\right)^{2/(n-1)}
\cos \left(\frac{\pi }{n-1}\right)\,,
\end{eqnarray}
while for $n=3$, the coefficient
$A_\gamma(n=3)=\pi^2$, and 
the integral for $A_\omega(n=3)$ does not converge. We refer
to Table~\ref{table1} for the values of $A(n)$ coefficients,
for the cases $n=4,5,6$.

%
%
\begin{table}[t!]
\setlength{\tabcolsep}{25pt}
\caption{\label{table1} Coefficients $A_\omega(n)$ and
$A_\gamma(n)$ for $n=4,5,6$.}
\begin{indented}
\item[]\begin{tabular}{@{}lll}
\br
$n$ & $A_\omega(n)$ & $A_\gamma(n)$ \\
\mr
4 & 9.84895 & 5.68629 \\
5 & 4.54652 & 4.54652 \\
6 & 2.93624 & 4.04139 \\
\br
\end{tabular}
\end{indented}
\end{table}

%
%
\begin{table*}[t!]
\caption{\label{table2}
Coefficients of collisional broadening and frequency 
shift are given for hydrogen transitions 
of experimental interest for high-precision spectroscopy.
The corresponding cross-sections can be calculated 
using the given data, via formula~(\ref{sigmaOmega6V}). 
The colliding system is described in the column ``perturber-spectator'' in the format 
``state of perturber atom--$($lower state of spectator 
atom$-$upper state of spectator atom$)$''.
The $1S$ hydrogen atoms may be present in all possible hyperfine substates, 
while the $2S$ atoms,
in accordance with the experimental apparatus,
are assumed to be in an $F=0$ substate only,
after having been excited via two-photon absorption from the 
$F=0$ hyperfine ground-state sublevel. 
This is at variance with Eqs.~(\ref{4P12_1S}),~(\ref{4P32_1S}),%
~(\ref{6P12_1S}), and~(\ref{6P32_1S}).
The coefficients $\xi^{(6)}_{\omega,\gamma}$ 
for perturber $1S$ atoms were averaged over the manifold of all available hyperfine 
substates, while the $2S$ perturber atoms were taken only in the $F=0$ substates. 
The averaging is done according to Eq.~(\ref{C6:2by5:Avg}).}
\begin{indented}
\item[]\begin{tabular}{l S[table-format=-1.3e-2] S[table-format=-1.3e-2]}
\br
Perturber-Spectator  
& \multicolumn{1}{c}{$\xi^{(6)}_\omega$\,[rad m$^2$ $($m/s$)^{2/5}$]}
& \multicolumn{1}{c}{$\xi^{(6)}_\gamma$\,[rad m$^2$ $($m/s$)^{2/5}$]}
\\ 
\mr
$1S$--$(1S-2S)$                &-2.232e-17 & 3.072e-17 \\
$1S$--$(1S-3S)$                &-4.325e-17 & 5.953e-17 \\
$1S$--$(1S-4S)$		       &-6.855e-17 & 9.435e-17 \\
$1S$--$(2S-4P_{1/2})$	       & 3.133e-16 & 4.313e-16 \\
$1S$--$(2S-4P_{3/2})$	       &-5.753e-16 & 7.919e-16 \\
$1S$--$(2S-6P_{1/2})$	       & 1.506e-16 & 2.072e-16 \\
$1S$--$(2S-6P_{3/2})$	       &-3.172e-16 & 4.365e-16 \\
$2S(F=0)$--$(1S-2S)$ 	       &-1.474e-15 & 2.029e-15 \\
$2S(F=0)$--$(2S-4P_{1/2})$     & 2.719e-14 & 3.742e-14 \\
$2S(F=0)$--$(2S-4P_{3/2})$     &-1.812e-14 & 2.494e-14 \\
$2S(F=0)$--$(2S-6P_{1/2})$     & 5.053e-15 & 1.304e-14 \\
$2S(F=0)$--$(2S-6P_{3/2})$     &-4.355e-14 & 5.967e-14 \\
\br
\end{tabular}
\end{indented}
\end{table*}

In case of collisions between hydrogen atoms, 
when the interaction potential is caused by van der Waals forces, 
the most relevant distance region is ($a_0 \ll R \ll a_0/\alpha$).
In the van der Waals region, the dipole interaction of an excited atom interacting
with the induced dipole of the  ground state atom produces van der Waals type of 
pressure shift and  broadening. The corresponding  
cross-sections, for $ n=6$, are given by 
\begin{eqnarray}
\label{sigmaOmega6V}
\sigma^{(6)}_\omega (v) = \xi^{(6)}_\omega v^{-2/5}\,, 
\qquad
\sigma^{(6)}_\gamma (v) = \xi^{(6)}_\gamma v^{-2/5},
\end{eqnarray}
where the superscript in $C_6$ indicates that the interactions are 
of the nonretarded van der Waals type and the  
proportionality coefficients $\xi^{(6)}_{\omega,\,\gamma}$ are given by 
\numparts
\begin{eqnarray}
\label{xi:omega}
\xi^{(6)}_{\omega}&=&-\frac{3^{2/5}\,\sqrt{5-\sqrt{5}\,}}{2^{27/10}}\, \pi^{9/10}\,
\Gamma\left(\frac{3}{5}\right) \mathrm{sgn}(C_6)\;
\left( \frac{|C_6|}{\hbar} \right)^{2/5}
\nonumber\\
&=&-2.93624\; \mathrm{sgn}(C_6)\;\left( \frac{| C_6 |}{\hbar} \right)^{2/5}\,,\\
\label{xi:gamma}
\xi^{(6)}_{\gamma}
&=&-\frac{3^{2/5} \, \left(\sqrt{5}+1\right)}{2^{11/5}\times 5}\, \pi^{7/5}
\Gamma\left(-\frac{2}{5}\right) \left( \frac{| C_6 |}{\hbar} \right)^{2/5} 
\nonumber\\
&=& 4.04139\;\left( \frac{| C_6 |}{\hbar} \right)^{2/5}\,.
\end{eqnarray}
\endnumparts
The coefficients $\xi^{(6)}_{\omega,\,\gamma}$,
computed for the different 
transitions are given in Table \ref{table2}. 
Numerically, the magnitude of the pressure-shift cross-section is 
about three-quarters of the pressure-broadening cross-section,
as is evident from
Eqs.~(\ref{xi:omega}) and (\ref{xi:gamma}). 

We refer to Refs.~\cite{AdEtAl2017vdWi,JeEtAl2017vdWii,%
JeAdDaMaKo2018jpb1,AdEtAl2017vdWiii,JeAd2017atoms} 
for numerical values of the van der Waals $C_6$ coefficients. 
As is evident from the discussion in Sec.~\ref{sec31},
the van der Waals $C_6$ coefficients 
are symmetry dependent 
for $nS$--$1S$ interactions
(see Refs.~\cite{AdEtAl2017vdWi,AdEtAl2017vdWiii,Ch1972}).
With $D_6$ denoting the ``direct'' term and $M_6$ the ``mixing'' term,
one obtains the coefficients $D_6 \pm M_6$ for the 
symmetric and anti-symmetric combinations of the 
two-atom states. The coefficients $D_6 \pm M_6$ should then
be taken to the power of $2/5$.
Assuming an equal likelihood for collisions to take 
place in either of the two symmetries, 
appropriate formulas to evaluate the $\xi^{(6)}_{\omega}$ 
and  $\xi^{(6)}_\gamma$ are given as 
\numparts
\begin{eqnarray}
\label{xi:omega:SD}
\fl \xi^{(6)}_{\omega}=-1.46812
\left\{ \left[\mathrm{sgn}(D_6 + M_6)\left( \frac{| D_6 + M_6 |}{\hbar} \right)^{2/5} 
+\mathrm{sgn}(D_6 - M_6)\left( \frac{| D_6 - M_6 |}{\hbar} \right)^{2/5}\right]\right.
\nonumber\\
\fl \qquad \pm\,
\left.\left[ \mathrm{sgn}(D_6 + M_6)\left( \frac{| D_6 + M_6 |}{\hbar} \right)^{2/5} 
-\mathrm{sgn}(D_6 - M_6)\left( \frac{| D_6 - M_6 |}{\hbar} \right)^{2/5}\right]\right\}\,,\\
\label{xi:gamma:SD}
\fl\xi^{(6)}_{\gamma}= 2.02070\left\{ \left[ \left( \frac{| D_6 + M_6 |}{\hbar} \right)^{2/5} 
+\left( \frac{| D_6 - M_6 |}{\hbar} \right)^{2/5}\right]\right.
\nonumber\\
\fl \qquad \pm\; 
\left.\left[\left( \frac{| D_6 + M_6 |}{\hbar} \right)^{2/5} 
-\left( \frac{| D_6 - M_6 |}{\hbar} \right)^{2/5}\right]\right\} \,.
\end{eqnarray}
\endnumparts
For all $nS$--$1S$ systems with $n \geq 4$, the mixing van der Waals coefficient 
$M_6$ is smaller by at least four order of magnitude than the 
direct term $D_6$~\cite{AdEtAl2017vdWiii}.
This fact simplifies the situation.
If we include the mixing term,
then, for the $2S$--$1S$ system, 
$\xi^{(6)}_\omega (v)$ and $\xi^{(6)}_\gamma (v)$ should read as 
\begin{eqnarray}
\label{data1}
\xi^{(6)}_{\omega}(2S;1S) = -(2.232\pm 0.142) \times 10^{-17}\,
\mbox{rad m$^2$ (m/s)$^{2/5}$} \,,
\end{eqnarray}
and
\begin{eqnarray}
\label{data2}
\xi^{(6)}_{\gamma}(2S;1S) = (3.072\pm 0.196) \times 10^{-17}\,
\mbox{rad m$^2$ (m/s)$^{2/5}$} \,,
\end{eqnarray}
respectively. The mixing term is not indicated
in Table~\ref{table2}, but given in 
explicit form in Eqs.~(\ref{data1}) and~(\ref{data2}). 
For the $3S$--$1S$ system, one has
\begin{eqnarray}
\xi^{(6)}_{\omega}(3S;1S) = -(4.3253\pm 0.0056) \times 10^{-17}\,
\mbox{rad m$^2$ (m/s)$^{2/5}$} \,,
\end{eqnarray}
and
\begin{eqnarray}
\xi^{(6)}_{\gamma}(3S;1S) = (5.9534\pm0.0078)\times 10^{-17}\,
\mbox{rad m$^2$ (m/s)$^{2/5}$} \,.
\end{eqnarray}
The mixing contribution only enters at the third decimal.
The long-range interaction
potential is attractive, which means that the frequency 
shift is negative. Presenting our results only to four significant figures, 
we can neglect mixing contributions for the $4S$--$1S$ 
system and higher excited reference states. 

In Table~\ref{table2}, we present the average values of $\xi^{(6)}_{\gamma}$ 
and   $\xi^{(6)}_{\omega}$ for $F=0$ to $F=1$ transitions 
in the hyperfine manifolds of the $2S$--$2S$ and $2S$--$nP_J$ systems.  
Note that both $\xi^{(6)}_{\gamma}$ 
and   $\xi^{(6)}_{\omega}$ are proportional to $| C_6 |^{2/5}$. 
The averaging over the fine 
and hyperfine levels is done as follows. 
Let us assume that the system we are interested in
has  $N$ hyperfine manifolds which we label by a subscript $j = 1, \ldots, N$,
with multiplicities $\mathbbm{m}_j$,
and also let $\mathbbm{M} = \sum_j \mathbbm{m}_j $.
The $\xi^{(6)}_{\gamma}$ and   $\xi^{(6)}_{\omega}$ are calculated using  
\numparts
\begin{eqnarray}
\xi^{(6)}_{\omega} &=& -2.93624\; 
\mathrm{sgn}(C_6)\;\langle | C_6 |^{2/5} \rangle/\hbar^{2/5} \,,
\\[0.1133ex]
\xi^{(6)}_{\gamma} &=& 4.04139\;\,\langle | C_6 |^{2/5} \rangle/\hbar^{2/5}\,,
\end{eqnarray}
\endnumparts
where
\begin{eqnarray}\label{C6:2by5:Avg}
\langle | C_6^{2/5} | \rangle &=&
\frac{1}{\mathbbm{M}} \,
\sum_j \mathbbm{m}_j \, \left( | C_6^{(j)} | \right)^{2/5} \,.
\end{eqnarray}
We shall notice that in our theoretical works 
(Refs.~\cite{JeAdDaMaKo2018jpb1, JeAd2017atoms}),
the energy shift for both the $1S$ and $2S$ collisions 
are averaged over all possible hyperfine states. 
The particular interest of the current work is concentrated on
the  $2S$--$4P$ and $2S$--$6P$ experiments at
the Max Planck Institute at Garching, where atoms are prepared in 
the metastable $2S(F=0)$ state~\cite{BeEtAl2017}. 
Selection rules of spectroscopy allow the excitation from 
the $2S(F=0)$ state to the $nP_{j}(F=1)$ states, 
where $j=1/2,\, 3/2$. So, the averaging in this case 
should be done over a different manifold of quantum states. 
We here use the van der Waals coefficients $C_6(2S(F=0)$--$nP_j(F=1))$
whose evaluation is described in detail in 
the accompanying article~\cite{JeAdDaMaKo2018jpb1}.

One may notice that in $nS-mP$ collisions,
the van der Waals Hamiltonian (\ref{vdw}) 
mixes the quantum states of the system $|nS_A,mP_B\rangle$ with a state 
$|mP_A,nS_B\rangle$.
In principle, this mixing may cause a first-order energy shift, proportional to $R^{-3}$.
However, this first-order shift averages out to zero 
when taken over state manifolds;
hence it does not contribute to the pressure shift~\cite{AlGr1965}.  
Indeed, we may cite the following remark on p.~1045 
of Ref.~\cite{AlGr1965}:
``Shifts would be given by the imaginary part
of the $S$ matrix element, but are zero for
resonance dipole-dipole interactions". 
Note that in the notation of Ref.~\cite{AlGr1965},
the $S$ matrix element is given by the expression~$\langle l | \Phi | l \rangle$.
In our notation, in~\ref{appA}, 
the $S$ matrix element is denoted as 
$\langle l | \theta | l \rangle$ in order to ensure the 
self-consistency of the 
notation we are using in this paper
(more details are provided in~\ref{appA}).

%
%
\section{Collisional shift in the Garching 
\texorpdfstring{$\bm{2S}$--$\bm{4P}$}{2S--4P} hydrogen experiment}
\label{sec4}

%
%
\subsection{Experimental apparatus} 
\label{sec41}

\begin{figure}[t!]
\begin{center}
\includegraphics [width=0.8\columnwidth]{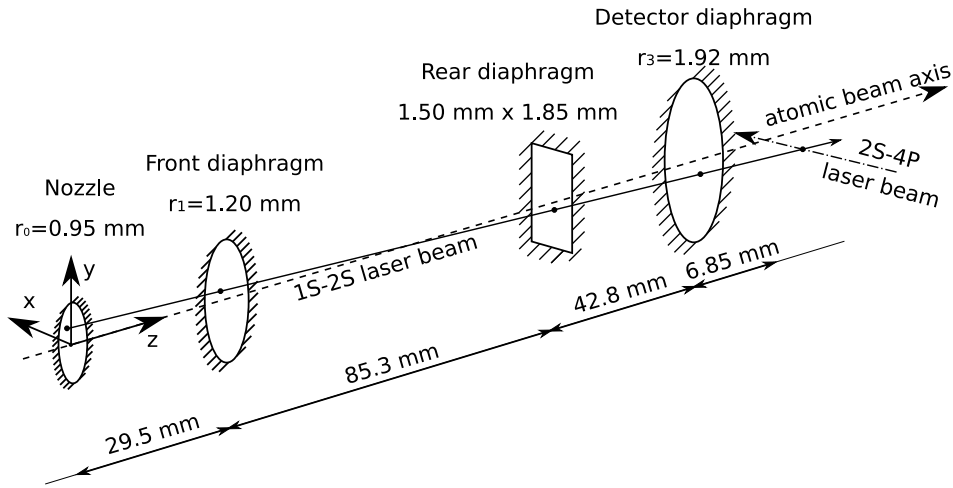}
\caption{
The scheme used for simulation of the collisional shift from intra-beam collisions. 
The trajectories of atoms are starting on the nozzle outlet;
the diaphragms are forming the atomic beam. 
Atoms can be excited in $2S$ state in the $1S$--$2S$ Gaussian laser beam 
with $1/e^2$ intensity radius $0.3$ mm. Excited $2S$ atoms 
crosses the $2S$--$4P$ spectroscopy laser beam. 
During that the $2S$--$4P$ beam the atoms can 
collide with other atoms from the beam, which causes an 
intra-beam shift. The Monte-Carlo procedure of calculation of this 
shift is based on random seeding of the trajectory of 
the spectator atom $A$, the trajectory of the perturber atom $B$ 
in a way that they collide in point $\vec{r}_{\rm col}$, 
averaging a function $\Omega_{sv}$, describing the collisional 
shift for different velocity groups with proper weights.}
\label{fig2}
\end{center}
\end{figure}

In order to evaluate the effect of the collisions between the atoms inside
the beam, we need to consider the geometry of the experimental setup.  A detailed
description of the $2S$--$4P$ experiment in Garching is given in
Refs.~\cite{BeEtAl2017, BeEtAl2015}. 
For the calculation of the collisional shift, we use a simplified model, 
based on the actual geometry of the experiment (see Fig.~\ref{fig2}).
According to this model, hydrogen atoms in the $1S$ state are emitted from a
cryogenic circular nozzle with a temperature 
of $T=5.8$\,K. The atomic beam is collimated by
several diaphragms, aligned along the $1S$--$2S$ 
two-photon excitation laser beam at $243$\,nm. 
After passing through the excitation region, where they can be excited to
the $2S$ state, the
atoms come into the $2S$--$4P$ spectroscopy region, 
in which they cross the $486$ nm laser beam. A possible 
bias due to simultaneous irradiation of the $2S$ atoms by 
the $1S$--$2S$ laser beam during the $2S$--$4P$ spectroscopy is avoided by 
shuttering the $1S$--$2S$ laser beam with rate of $160$\,Hz. 
The $2S$--$4P$ excitation signal is detected via Ly-$\alpha$ 
and Ly-$\gamma$ decays and recorded only in the periods when the $1S$--$2S$ 
laser beam is blocked.  A multi-channel scaler separates the signal from the 
$2S$--$4P$ spectroscopy into several groups according
to the delay between the beam blocking falling
front and the photon detection.  The velocity distribution of the
atoms  $v_A$ strongly depends on the delay,
since fast atoms pass the $2S$--$4P$ spectroscopy region earlier than slow ones.

The total flux of hydrogen atoms in our calculation is taken 
from the measured flow of the hydrogen gas into the apparatus during the 
$2S$--$4P$  measurement (approximately $1.8\times 10^{18}$ H$_2$ molecules
per second). 
An estimate of the dissociation rate into hydrogen atoms
can be obtained indirectly, via two independent methods, 
which independently indicate a fraction of atomic hydrogen on the order
of~$10$\%. The first estimate proceeds as follows.
Upon a decrease in temperature from $20$\,K to $5.8$\,K, the residual
pressure near the nozzle is observed to decrease by more than 
a factor $10$. Within a crude approximation, we can
assume that molecular hydrogen, as opposed to atomic hydrogen,
becomes solid in this temperature range
and is left on the nozzle as ice, which means 
that about $90$\% of the gas remains in molecular form. 
Alternatively, we experimentally study 
how the amplitude $A(T)$ of the line changes as a function of temperature.
Within a crude approximation, one would 
have $A(T) = N \, d(T) \, e(T)$, where $N$ is the number of 
molecules, $d(T)$ is the dissociation ratio, 
and $e(T)$ is the excitation probability. If $N$ is independent 
of temperature,  and $e(T)$ is proportional to $1/T$
(see Ref.~\cite{Su1962}), then the amplitudes of the 
lines at $5.8$\,K and $300$\,K should be 
related to the corresponding dissociation rates as 
$A(5.8\,{\rm K})/A(300\,{\rm K})  \approx 
[ d(5.8\,{\rm K})/d(300\,{\rm K}) ] \, [ (5.8\,{\rm K})/(300\,{\rm K}) ]$. 
This estimate also is compatible with a dissociation rate 
not exceeding 10\,\%,
and a flow of hydrogen atoms thus not exceeding $3.6\times10^{17}$ atoms/s.

%
%
\subsection{Analytic estimate for intra--beam collisions} 
\label{sec42}

It is interesting to note that the collisional shift can be estimated via a simple 
analytic calculation. We notice that the average velocity of the atoms 
leaving the nozzle in the case of a Maxwell distribution with temperature 
$T=5.8$\,K is about $v= 3 \sqrt{\pi/8 \times k_B T/m_H}\approx 410$ m/s. 
If  all the atoms are flying from a point-like source with a
uniform angular distribution, the number of atoms crossing the sphere of 
radius $L_1$ can be estimated as $N=4\pi \, L_1^2 \, v \, \mathbbm{n} $, where 
$\mathbbm{n}$ is a number density of the atoms. 
From the flux $N=3.6\times10^{17}$ atoms/s,
one can estimate the concentration of ground-state hydrogen atoms in the atomic beam 
at a distance of about $L_1=16.4$\,cm from the nozzle to 
be about $\mathbbm{n} = N/(4\pi \, L_1^2 \,v)\approx 2.6\times10^{15}$ atoms/m$^{3}$. 
For $1S$--$4P_{3/2}$ collisions with a velocity of $410$ m/s, the collisional shift 
cross-section is about 
$\sigma_\omega = -5.1\times 10^{-17}$ rad m$^2$, 
which finally gives a shift $\omega_c/(2\pi) = \mathbbm{n} \, v \,
\sigma_\omega/(2\pi)\approx -8.6 $ Hz. 
We recall that for $4P_{1/2}$, the van der Waals potential is repelling, 
while for $4P_{3/2}$, it is attractive~\cite{JeAdDaMaKo2018jpb1}.
This result is close to the result of 
the Monte-Carlo simulations, confirming the possibility to neglect  
pressure-related effects on the current level of experimental 
uncertainty~\cite{BeEtAl2017, BeEtAl2015}.

%
%
\subsection{Monte--Carlo calculation for intra--beam collisions} 
\label{sec43}

The simple analytic estimate
given above can be criticized since we ignore the complicated 
spatial and velocity distributions of the hydrogen atoms. 
In order to take this into account, 
we use an approach based on a Monte-Carlo simulation. In the framework of this simulation, 
we consider the collisional shift caused by collisions with other atoms from the same 
beam. These can be in either the $1S$ or $2S$ state.  
For simplicity, we do not model a full lineshape of the $2S$--$4P$ spectroscopy, 
but restrict our evaluation to the collisional shift,
equally weighted over the set of $2S$ atoms. 
For the computation of the excitation probability of the 
spectator atoms $A$ in the delay window from,
say, $\tau_1$ to $\tau_2$, 
we use an existing Monte-Carlo approach, described in 
Refs.~\cite{HaEtAl2006,Ha2006phd}. 
The origin points of trajectories of the hydrogen atoms  
in this simulation are seeded uniformly 
on the orifice of the nozzle, while the velocity vectors of these trajectories 
are seeded uniformly in the corresponding solid angle. 
The trajectories which do not pass all  diaphragms are rejected.
The program is also choosing a random absolute value of 
the velocity $v_A$ of the atom 
according to a Maxwellian distribution, and the random position 
of the atom $z_{\rm off}$ on the $z$-axis, where the atom is 
located at the moment of shuttering the excitation light. The Monte-Carlo procedure 
discards the seeded trajectory if the atom does not fall into the delay window,
characterized by the condition $\tau_1<\tau<\tau_2$. Here, we designate 
as $\tau=(L_0-z_{\rm off})/v_{Az}$ an individual delay of the atom $A$, 
where $L_0$ is the distance from the nozzle to the $2S$--$4P$ laser beam, 
and $v_{Az}$ is the $z$ component of the velocity of the atom.  
For all the atoms which fit into the delay window, we can compute the excitation 
probability $\rho_{22}$ to the $2S$ state by integrating the
optical Bloch equations from the moment of the 
beginning of trajectory to the moment when the light is shuttered off.

The result of the procedure described above is a set of random atomic trajectories 
described by the initial position $\vec{r}_{A0}$, the velocity vector $\vec{v}_A$ 
and the individual delay of the atom $\tau$.  For each trajectory, one computes 
the excitation probability to the $2S$ state $\rho_{22}$, and the position of the atom $\vec{r}_{\rm col}$, 
when it is crossing the $2S$--$4P$ laser beam. Random 
collisions with other atoms $B$ in the vicinity of 
the point $\vec{r}_{\rm col}$ cause the intra-beam shift,
where we assume that the modulus of $\vec{r}_{\rm col}$ is 
large against the distance of closest approach of the 
two atoms during the collision, i.e.,
large against the impact parameter $b$.
This means that $| \vec{r}_{\rm col} | \gg | \vec r_A(t_{\rm col})  - \vec r_B(t_{\rm col}) | \equiv b$,
where $t_{\rm col}$ is the point in time of closest approach of the 
two atoms. The weight for this averaging is the excitation probability 
of the atom~$A$, which we denote as $\rho_{22}$.
 
The evaluation of the collisional shift of the atom $A$ in the
position $\vec{r}_{\rm col}$ with velocity $\vec{v}_A$ can be done via 
the cross-sections of the collisional shifts, computed by formula (\ref{sigmaOmega6V}). 
In order to evaluate collision rates and velocities, we can describe our nozzle as a set of point-like 
sources of atoms $B$. Each source $s$ with position $\vec{r}_s$ emits $N_s$ 
atoms per second with a velocity distribution $p(v_B)$ and an
angular distribution $\Psi(\hat{e})$, where $\hat{e}$ is a unit vector representing direction. 
We assume that atomic trajectories are straight during the $1S$--$2S$ 
excitation phase, so that $\hat{e}$ can be expressed as
\begin{eqnarray}
\hat{e} = \frac{\vec{r}_{\rm col}-\vec{r}_s}{|\vec{r}_{\rm col}-\vec{r}_s|} .
\end{eqnarray}

For most atoms contributing to the experimentally 
observed line shape,
the assumption of a straight trajectory of the atoms 
during the $1S$--$2S$ excitation (which is of course different from the 
subsequent $2S$--$4P$ excitation) is valid since the probability of 
the collision with a small deflection angle (which allows the atom 
to pass through the front and rear diaphragms) is quite small. 
For example, our estimation shows, that the fraction of atoms, 
which can be deflected by the angle between 
$10^{-4}$ and $10^{-1}$ rad, is less than $10^{-3}$. 
In our experimental geometry, that implies that either,
the atoms in the atomic beam in the $2S$--$4P$ spectroscopy region 
come to this region from the nozzle without a strong deviation from 
the straight-line trajectory, or they do not get there at all. 

A remark is in order, which is relevant especially for the 
$2S$--$4P$ excitation region, because of the 
large $C_6$ coefficients governing atoms
in the excited $P$ state. As we show in~\ref{appC}, it is interesting to 
observe that the assumption that the trajectory of the atom 
is close to a straight line during the collision, 
actually is not fulfilled in the complete range of impact parameters
relevant to our calculation. When the impact parameter is close to 
a ``deflection radius'', which for our geometry is close to the 
Weisskopf radius,  then the trajectories of the colliding particles are strongly deflected. 
In this case, the problem of the collisional shift should be considered 
with a full account of the experimental geometry.
In the setup of the Garching $2S$--$4P$ experiment,
the atoms, deflected in the region of $2S$--$4P$ spectroscopy, 
remain in a spatial region where the emitted 
decay photons could be registered by the detector,
upon decay,
in the form of Lyman-$\alpha$ (after quenching) 
and Lyman-$\gamma$ photons (from the $4P$ state).

Experimentally, if the $2S$ atom is kicked out of the beam, 
then it will hit the wall of the detector ``box''. 
With a high probability, a  collision with the wall leads to a quenching of the $2S$ state, 
due to the interaction  of the atom with the surface of the grounded conductor,
and emission of a Lyman-$\alpha$ photon. 
(In our experiment, 
we actually do not detect the photons, but the photoelectrons, which they kick out from the walls,
as described in Ref.~\cite{BeEtAl2017}.)

When $2S$ (or, conceivably, $4P$) atoms are kicked out of the beam
by collisions with small impact parameters, they do so irrespective of the 
frequency of the spectroscopy laser.
In other words, if the $2S$ atom leads to an event registered 
by the detector,  regardless of the state of $2S$--$4P$ laser, 
the atom could only contribute to the constant background, 
which is eliminated by our fitting procedure~\cite{BeEtAl2017}. 
Only the atoms with a small deflection angle 
(less than $10^{-1}$~rad) contribute to the observed 
resonance line shape. The latter escape from the detector undetected 
if the $2S$--$4P$ laser is off-resonance and are detected via decay 
from the $4P$ state if the $2S$--$4P$ laser is in resonance. 
In our Monte Carlo simulation, we assume that 
the trajectories of the colliding atoms are straight, 
and ignore the possibility of a large deflection from the 
straight-line trajectory. Thus,
we very likely overestimate the observed collisional shift, 
because we also take into consideration the atoms with a large deflection angle;
these should otherwise be ignored because they only contribute to 
the background. In consequence, 
a Monte Carlo simulation with 
straight trajectories during the $2S$--$4P$ excitation can be used as an estimate of 
an upper limit of the collisional shift. 

The number density of atoms $B$ with velocities in the interval $(v_B,v_B+ {\rm d}v_B)$, 
created by the source $s$ in the point $\vec{r}_{\rm col}$, can be evaluated as
\begin{eqnarray}
\dd n_{sv} = \frac{N_s}{4 \pi v_B |\vec{r}_{\rm col}-\vec{r}_s|^2} 
\Psi\left(\frac{\vec{r}_{\rm col}-\vec{r}_s}{|\vec{r}_{\rm col}-\vec{r}_s|} \right) \,
\Xi(\vec{r}_s,\vec{r}_{\rm col}) \, p(v_B)  \, \dd v_B,
\end{eqnarray}
where $\Xi(\vec{r}_s,\vec{r}_{\rm col})$ is a filtering function, 
which is equal to unity if the trajectory, 
starting at $\vec{r}_s$ and heading towards $\vec{r}_{\rm col}$, passes
all diaphragms installed in our experiment, 
and zero if not. The relative velocity of the atoms $A$ and $B$ is 
\begin{eqnarray}
\label{vcol}
\vec v_{\rm col}=\vec{v}_A-\vec{v}_B 
= \vec{v}_A-v_B \, \frac{\vec{r}_{\rm col}-\vec{r}_s}{|\vec{r}_{\rm col}-\vec{r}_s|} \,, 
\end{eqnarray}
and so, the infinitesimal frequency shift 
$\dd\omega_{sv}$ caused by the source $s$ and velocity component $(v_B, v_B + \dd v_B)$ is:
\begin{eqnarray}
\dd\omega_{sv} &= & \frac{N_s}{4 \pi v_B |\vec{r}_{\rm col}-\vec{r}_s|^2} \,
\Psi\left(\frac{\vec{r}_{\rm col}-\vec{r}_s}{|\vec{r}_{\rm col}-\vec{r}_s|} \right) 
\Xi(\vec{r}_s,\vec{r}_{\rm col}) \, p(v_B)  \, \dd v_B \nonumber\\
& & \times \left|\vec{v}_A - v_B \frac{\vec{r}_{\rm col}-\vec{r}_s}{|\vec{r}_{\rm col}-\vec{r}_s|}\right| 
\; \sigma_\omega \left(\left|\vec{v}_A - v_B 
\frac{\vec{r}_{\rm col}-\vec{r}_s}{|\vec{r}_{\rm col}-\vec{r}_s|}\right|\right) \,.
\end{eqnarray}

The angular distribution $\Psi$, for practial calculations, is taken 
proportional to the  $\cos(\theta) $, where the $\theta$ is the angle 
between the normal vector to the surface of the nozzle 
outlet and the atomic trajectory. This choice is motivated by
Lambert's cosine law, which is well known for ideal diffusive radiators. 
In order to write this probability distribution function explicitly, 
we can introduce another angle $\varphi$ so that $\theta$ and $\varphi$ 
define a spherical coordinate system, whose $z$ axis points away
from the point of the source in direction of the normal vector to the nozzle surface. 
In this coordinate system, the probability distribution $\Psi$ 
can be written as
%
\begin{equation}
\Psi(\theta,\varphi) =
\cos(\theta)\sin(\theta)/\pi,
\end{equation}
which is normalized to
\begin{equation}
\int_0^{2 \pi} \dd \varphi  \, \int_0^{\pi/2} \dd  \theta \,
\Psi(\theta,\varphi) = 1\,.
\end{equation}
Indeed, for our nozzle, the relevant angular ranges are
$0<\theta<\pi/2$ and $0<\varphi<2\pi$.
\color{black}

\begin{table*}[t]
\caption{\label{table3}  
Results are given for a Monte-Carlo evaluation of 
the total intra-beam collisional 
shift $f(X) = \omega_{\rm col}$, 
as defined in Eq.~(\ref{omega_total}),
where $X$ is a delay interval, as discussed in the text.
The data are relevant for the 
$2S$--$4P$ experiment, and the corresponding $C_6$ coefficients. 
The $C_6$ coefficient is given in atomic units
(i.e., in units of $E_h \, a_0^6$),
while the frequency shifts are given in Hz.
The notation $f(\tau_1-\tau_2) = \omega_{\rm col}(\tau_1-\tau_2)/(2\pi)$ 
denotes a frequency shift in Hz, 
computed for the delay window from
$\tau_1$ to $\tau_2$, i.e., for atoms
that arrive within the given delay interval after the beam blocking.
The values of $\tau_i$ are given in $\mu$s. 
The uncertainty of the Monte-Carlo evaluation 
is on the order of 3\%.} 
\begin{indented}
\item[]
\begin{tabular}{l S[table-format=-1.3e0] S[table-format=-1.3e0] 
S[table-format=-1.3e0] S[table-format=-1.3e0] }
\br
System
& \multicolumn{1}{c}{$1S$--$(2S$--$4P_{1/2})$}
& \multicolumn{1}{c}{$2S$--$(2S$--$4P_{1/2})$} 
& \multicolumn{1}{c}{$1S$--$(2S$--$4P_{3/2})$}
& \multicolumn{1}{c}{$2S$--$(2S$--$4P_{3/2})$} \\ 
\mr
$C_6$ [a.u.]
& \multicolumn{1}{c}{$1.296 \times10^5$} 
& \multicolumn{1}{c}{$9.090 \times10^9$}
& \multicolumn{1}{c}{$-5.921\times10^5$} 
& \multicolumn{1}{c}{$3.296 \times10^9$} \\
\mr
$f(10-60)$ [Hz] & 5.4 & 0.92 & -10.0& 0.61 \\ 
$f(60-110)$ [Hz] & 5.3 & 0.75& -9.8 & 0.50 \\ 
$f(110-160)$ [Hz] & 5.1 & 0.61 & -9.3 & 0.41 \\ 
$f(160-210)$ [Hz] & 4.7 & 0.48 & -8.7 & 0.32 \\ 
$f(210-260)$ [Hz] & 4.6 & 0.35 & -8.5 & 0.24 \\ 
$f(260-310)$ [Hz] & 4.5 & 0.25 & -8.2 & 0.17 \\ 
$f(310-410)$ [Hz] & 4.5 & 0.17 & -8.3 & 0.11 \\ 
$f(410-610)$ [Hz] & 4.8 & 0.076 & -8.8 & 0.051 \\ 
$f(610-810)$ [Hz] & 5.2 & 0.024 & -9.6 & 0.016 \\ 
$f(810-2560)$ [Hz] & 6.1& 0.005 & -11.3 & 0.003\\
\br
\end{tabular}
\end{indented}
\end{table*}

In order to compute the collisional shift of the 
individual atom $A$, this function should be integrated over all the velocities 
$v_B$ and all the sources of atoms $s$. It is clear that any attempt at an
analytic computation of this integral 
would be hopeless even in 
simple cases. However, we can use the Monte-Carlo method with good effect.
It is advantageous to restrict possible values of the velocity $v_B$ 
to an interval from zero to $v_{\rm max}$,
where $v_{\rm max}$ is a velocity chosen to be bigger than the velocity of most of our atoms.
One then needs to calculate the average value of the function
\begin{eqnarray}
\Omega_{sv} & =  & 
\frac{1}{N_s} \, \frac{\dd \omega_{sv}}{\dd v_B} = 
\frac{1}{4 \pi v_B |\vec{r}_{\rm col}-\vec{r}_s|^2} 
\Psi\left(\frac{\vec{r}_{\rm col}-\vec{r}_s}{|\vec{r}_{\rm col}-\vec{r}_s|} \right)
 \Xi(\vec{r}_s,\vec{r}_{\rm col}) \, p(v_B) \nonumber\\
& & \times 
\left|\vec{v}_A - v_B \frac{\vec{r}_{\rm col}-\vec{r}_s}{|\vec{r}_{\rm col}-\vec{r}_s|} \right| 
\; \sigma_\omega \left(\left|\vec{v}_A - v_B 
\frac{\vec{r}_{\rm col}-\vec{r}_s}{|\vec{r}_{\rm col}-\vec{r}_s|}\right|\right).
\end{eqnarray}
For this averaging, we should seed the velocity of the atom $v_B$ according to a
uniform distribution, since the function $\Omega_{sv}$ contains the 
properly normalized probability density function $p(v_B)$. 
The sources of the atom can be seeded with a weight
determined according to their flux $N_s$, which practically means 
that in each Monte-Carlo run, we can randomly choose, as the point 
of origin of atom $B$, a 
specific point $\vec{r}_{s}$ on the nozzle orifice. The total shift can be evaluated as:
\begin{eqnarray}
\label{omega_total}
\omega_{\rm col}=\sum_s \int\limits_0^{v_{\rm max}} \dd \omega_{sv} 
\approx N_{\rm tot} \; v_{\rm max} \; \langle \Omega_{sv} \rangle_{\rm MC},
\end{eqnarray}
where $N_{\rm tot}$ is a sum of fluxes of all the sources, 
and $\langle \cdots \rangle_{\rm MC}$ represents 
a Monte-Carlo averaging. Notice, that in this approach to the 
intra-beam collisional shift, we
do not necessarily need to use discrete sources of the perturbing atoms. 
The sources can be distributed on some surface or even in an extended volume. 
For the purpose of our evaluation, we assume that sources are uniformly 
distributed on the orifice of the nozzle.

\begin{table}[t!]
\caption{\label{table4} 
Results are given for 
a Monte-Carlo evaluation of the intra-beam collisional shift 
$f(X) = \omega_{\rm col}$,
as defined in Eq.~(\ref{omega_total}),
with $X$ denoting a delay interval.
Here, we consider the 
$2S$--$6P$ experiment and the corresponding $C_6$ coefficients. 
The $C_6$ coefficients are given in atomic units 
and the frequency shifts in Hz. The notation 
$f(\tau_1-\tau_2) = \omega_{\rm col}(\tau_1-\tau_2)/(2\pi)$ 
means a frequency shift in Hz, computed for the delay window from 
$\tau_1$ to $\tau_2$, i.e., for atoms 
that arrive within the given delay interval after the 
beam blocking. The values of $\tau_i$ are given in $\mu$s. 
The uncertainty of the Monte-Carlo evaluation is on the order of 3\%.}
\begin{indented}
\item[]
\begin{tabular}{l S[table-format=-1.3e0] S[table-format=-1.3e0] 
S[table-format=-1.3e0] S[table-format=-1.3e0] }
\br
System
& \multicolumn{1}{c}{$1S$--$(2S$--$6P_{1/2})$}
& \multicolumn{1}{c}{$2S$--$(2S$--$6P_{1/2})$}
& \multicolumn{1}{c}{$1S$--$(2S$--$6P_{3/2})$}
& \multicolumn{1}{c}{$2S$--$(2S$--$6P_{3/2})$} \\ 
\mr
$C_6$ [a.u.]
& \multicolumn{1}{c}{$2.074\times10^4$} 
& \multicolumn{1}{c}{$4.280\times10^{10}$} 
& \multicolumn{1}{c}{$-1.336\times10^5$} 
& \multicolumn{1}{c}{$2.918\times10^{10}$} \\ 
\br 
$f(10-60)$ [Hz]     & 2.6 & 1.7   & -5.5 & 1.5 \\ 
$f(60-110)$ [Hz]    & 2.6 & 1.4   & -5.4 & 1.2 \\ 
$f(110-160)$ [Hz]   & 2.4 & 1.1   & -5.2 & 0.99\\ 
$f(160-210)$ [Hz]   & 2.3 & 0.89  & -4.9 & 0.76 \\ 
$f(210-260)$ [Hz]   & 2.2 & 0.66  & -4.6& 0.56 \\ 
$f(260-310)$ [Hz]   & 2.2 & 0.48  & -4.6 & 0.41\\ 
$f(310-410)$ [Hz]   & 2.2 & 0.31  & -4.6 & 0.26\\ 
$f(410-610)$ [Hz]   & 2.3 & 0.13  & -4.9& 0.12\\ 
$f(610-810)$ [Hz]   & 2.5 & 0.05  & -5.3& 0.04\\ 
$f(810-2560)$ [Hz]  & 2.9 & 0.009 & -6.2& 0.008\\ 
\br
\end{tabular}
\end{indented}
\end{table}

A particular effect which should be considered in the $2S$--$4P$ experiment,
concerns the possibility to collide with another $2S$ atom in the beam. 
In order to take this effect into account, we compute the probability $\rho_{22B}$ 
of excitation of each atom $B$. We can use the fact that the 
individual delay of atom 
$B$ must exactly coincide with the individual delay $\tau$ of the atom $A$. 
If the seeded velocity of atom $B$ satisfies the condition 
$v_{Bz} > L_0/\tau$, then atom $B$ leaves the nozzle after the 
light was shuttered off, so this atom 
cannot be excited to the $2S$ state and $\rho_{22B}=0$. 
If $v_B < L_0/\tau$, then atom $B$ can be excited to the 
$2S$ state with a probability, which can be 
computed by an integration of the optical Bloch equations. We also need to take 
into account, that the $1S$--$2S$ laser beam excites only 
atoms in the $1S(F=0)$ state, since the 
transition $1S(F=1) \rightarrow 2S(F=1)$ 
is not in resonance with the $1S$--$2S$ laser beam. 
According to statistical weight, the probability that 
the atom originates in a state $F=0$ is $1/4$, so we should multiply 
the result of the integration of the Bloch equations by the factor $1/4$. 
Thus, we can separate the collisional shift into two parts---collisions with $1S$ 
atoms and collisions with $2S$ atoms---by averaging two different functions:
\begin{eqnarray}
\Omega_{sv}^{1S} &=&  \frac{1-\rho_{22B}}{4 \pi v_B |\vec{r}_{\rm col}-\vec{r}_s|^2} 
\Psi\left(\frac{\vec{r}_{\rm col}-\vec{r}_s}{|\vec{r}_{\rm col}-\vec{r}_s|} \right) \;
\Xi(\vec{r}_s,\vec{r}_{\rm col}) \; p(v_B)
\nonumber\\
& & \times \left|\vec{v}_A - v_B \frac{\vec{r}_{\rm col}-\vec{r}_s}{|\vec{r}_{\rm col}-\vec{r}_s|}\right| 
\; \sigma_\omega^{1S}\left(\left|\vec{v}_A - v_B 
\frac{\vec{r}_{\rm col}-\vec{r}_s}{|\vec{r}_{\rm col}-\vec{r}_s|}\right|\right)\,,\\
\Omega_{sv}^{2S} &=&  \frac{\rho_{22B}}{4 \pi v_B |\vec{r}_{\rm col}-\vec{r}_s|^2} 
\Psi\left(\frac{\vec{r}_{\rm col}-\vec{r}_s}{|\vec{r}_{\rm col}-\vec{r}_s|} \right) \;
\Xi(\vec{r}_s,\vec{r}_{\rm col}) \; p(v_B) 
\nonumber\\
& & \times \left|\vec{v}_A - v_B \frac{\vec{r}_{\rm col}-\vec{r}_s}{|\vec{r}_{\rm col}-\vec{r}_s|}\right| 
\; \sigma_\omega^{2S}\left(\left|\vec{v}_A - v_B 
\frac{\vec{r}_{\rm col}-\vec{r}_s}{|\vec{r}_{\rm col}-\vec{r}_s|}\right|\right) \,.
\end{eqnarray}
The cross-sections of the shift $\sigma_\omega^{nS}$, with $n = 1, 2$,
can be computed using 
formula~(\ref{sigmaOmega6V}). The $C_6$ coefficient for $1S$ collisions 
should be averaged over all possible hyperfine sublevels of the $1S$ state,
while the $C_6$ coefficient for the $2S$ atoms should take 
into account only $2S$ atoms in the $F=0$ state. 
The $C_6$ coefficients used for our simulation together 
with the results of the calculation are given in Table~\ref{table3}.
It is instructive to perform a similar calculation, for the ongoing $2S$--$6P$ 
experiment in the same apparatus. Corresponding results are given in Table~\ref{table4}.

%
%
\subsection{Collisions with background gas}
\label{sec44}

Here, we consider the pressure shift caused by collisions with the background gas. 
A precise calculation of this shift has no practical interest because of the
unknown fraction of atomic hydrogen in the background gas, 
and fluctuations of the pressure in the $2S$--$4P$ spectroscopy 
region caused by cryopump working cycles. Thus, our goal is to 
conservatively estimate the shift from beam-background collisions, 
assuming that the effect from the molecules is 
less than or comparable to that caused by the atomic hydrogen
in the background gas. In fact, because of a much 
smaller $C_6$ coefficients for the 
atom-molecule as compared to the atom-atom
collisions (see~Sec.~5 of Ref.~\cite{JeAdDaMaKo2018jpb1}), this 
assumption should be well justified. 
The calculation described 
below takes only the effect of collisions with atomic hydrogen 
into account. In fact, we can easily do the calculation analytically 
under the assumption 
that the velocities of the atoms in the atomic beam are much smaller than 
the velocities of the background gas. 

The general setting is as follows. We consider an atomic beam, 
at a temperature of about $5.8$\,K, with hydrogen atoms inside the 
beam being perturbed by a $300$\,K background gas, consisting of 
atomic hydrogen in the ground state. The $1S$--$2S$ excitation probability 
depends on the velocity of the atom and the chosen experimental delay group, 
but for all delay groups, the average velocity of $2S$ atoms in the beam is 
less than $300$\,m/s. The thermal velocity of the background gas atoms is about 
$3$\,km/s, which means that we can neglect the movement of the spectator 
atom~$A$ in comparison with the perturber atom~$B$. Thus, the velocity of 
the collision is $v=|\vec{v}_A-\vec{v}_B|\approx v_B$. The pressure shift  
and the pressure broadening can be calculated from the known cross-sections: 
\begin{eqnarray}\label{sigmaC6Omega}
\omega_c  =\mathbbm{n}\int_0^\infty \sigma^{(6)}_\omega (v)\,v \,P(v)\, \mathrm{d}v\,,
\qquad 
\gamma_c  =\mathbbm{n}\int_0^\infty \sigma^{(6)}_\gamma (v)\,v \,P(v)\, \mathrm{d}v\,,
\end{eqnarray}
where  $P(v)$ is the velocity distribution of the background gas 
and $\mathbbm{n}$ is the number density of $1S$ atoms. 
For our estimation,
we can use a Maxwellian velocity distribution, which reads as follows,
\begin{eqnarray}
P(v)=\sqrt{\frac{2}{\pi}} \left(\frac{m}{k_B T} \right)^{3/2} v^2
\exp \left(-\frac{m v^2}{2 k_B T}\right),
\end{eqnarray}
where $m$ is a mass of background gas particle (hydrogen atom), $k_B$ is a Boltzmann constant, 
$T=300$\,K is a background gas temperature. 

Putting $\sigma^{(6)}_\omega (v) = \xi^{(6)}_\omega v^{-2/5}$ 
and  $\sigma^{(6)}_\gamma (v) = \xi^{(6)}_\gamma v^{-2/5}$,
the integral~(\ref{sigmaC6Omega}) can be computed analytically
with the result
\begin{eqnarray}
\label{ShiftFromXi}
\omega_c = \frac{2^{13/10}}{\sqrt{\pi}}\Gamma\left(\frac{9}{5}\right) \mathbbm{n}\, 
\xi^{(6)}_\omega \left(\frac{k_B T}{m}\right)^{3/10} 
=1.29388\times \mathbbm{n}\,\xi^{(6)}_\omega \left(\frac{k_B T}{m}\right)^{3/10}\,,\\
\label{BroadFromXi}
\gamma_c =1.29388\times \mathbbm{n}\,\xi^{(6)}_\gamma \left(\frac{k_B T}{m}\right)^{3/10}\,.
\end{eqnarray}
We recall that both the  $\xi^{(6)}_{\omega}$ and  $\xi^{(6)}_{\gamma}$ are 
proportional to $C_6^{2/5}$. Consequently,  both $\omega_c$ and $\gamma_c$ 
are also proportional to $C_6^{2/5}$.

At a temperature of $300$\,K and a pressure less than $10^{-8}$\,mbar,
the density of the background gas does not exceed $2.4\times 10^{14}$ m$^{-3}$. 
Under these conditions, the background shift in the $2S$--$nP$
experiments (with $n=4,6$) does not exceed the following values:
\numparts
\begin{eqnarray}
\label{numerics_a}
\fl\qquad \omega_c(1S-4P_{1/2})= 2 \pi \times 1.24\, \mathrm{Hz}\,,
\qquad  
\gamma_c(1S-4P_{1/2})= 2\pi \times 1.70\, \mathrm{Hz}\,, \\
\label{numerics_b}
\fl\qquad \omega_c(1S-4P_{3/2})=- 2\pi \times 2.34\, \mathrm{Hz}\,,
\qquad  
\gamma_c(1S-4P_{3/2})= 2\pi \times 3.22\, \mathrm{Hz}\,, \\
\label{numerics_c}
\fl\qquad \omega_c(1S-6P_{1/2})=2 \pi \times 0.57\, \mathrm{Hz}\,,
\qquad  
\gamma_c(1S-6P_{1/2})= 2\pi \times 0.78\, \mathrm{Hz}\,, \\
\label{numerics_d}
\fl\qquad \omega_c(1S-6P_{3/2})=- 2\pi \times 1.29\, \mathrm{Hz}\,,
\qquad  
\gamma_c(1S-6P_{3/2})= 2\pi \times 1.78\, \mathrm{Hz}\,. 
\end{eqnarray}
\endnumparts
The quantities $\omega_c$ and $\gamma_c$ in Eqs.~(\ref{numerics_a})--(\ref{numerics_d}) are 
the  hyperfine-structure averages of the shifts.
For both the frequency shift as well as the broadening,
the averaging scheme outlined in 
Eqs.~(\ref{C6:2by5:Avg}) has been used,
\begin{eqnarray}
 \omega_c\sim \left< |C_6|^{2/5}\right>\,, 
 \qquad
 \gamma_c\sim \left< |C_6|^{2/5}\right>\,.  
 \end{eqnarray}
The importance of the proper averaging procedure
is discussed in~\ref{appB}.

%
%
\section{Conclusions}
\label{sec5}

In this paper, we have outlined a procedure for the calculation of pressure
shifts in the Garching $2S$--$4P/6P$ experiments;
however, similar approaches can be used in other 
modern high-precision spectroscopic atomic-beam
experiments. The treatment is based on the impact
approximation (Sec.~\ref{sec2}), in which the phase and frequency
shifts in the collisions are modeled on the basis of
``quasi-instantaneous'' impacts onto the spectator atoms,
by colliding with perturber atoms. 
The basis for the calculation of the collisional 
shifts and broadenings is discussed in Sec.~\ref{sec3}.
An integration of the frequency shift, and of the pressure 
broadening, over the impact parameter $b$,
leads to results for the 
frequency-shift and broadening cross sections
which are proportional to $| C_6|^{2/5}$,
where $C_6$ is the van der Waals coefficient (see Sec.~\ref{sec33}).
The data in Table~\ref{table2}, with appropriate modifications
of the hyperfine 
averages~\cite{AdEtAl2017vdWi,JeEtAl2017vdWii,JeAdDaMaKo2018jpb1}, 
could be used for the description of pressure shifts 
in $1S$--$nS$ ($n=2,3,4$) and $2S$--$nP$ experiments ($n=4,6$).

An application of the developed formalism to 
recent and planned $2S$--$4P$ and $2S$--$6P$ experiments
is discussed in Sec.~\ref{sec4}.
After a discussion of the experimental apparatus in 
Sec.~\ref{sec41}, an analytic estimate of the 
frequency shift is presented in Sec.~\ref{sec42},
and a more elaborate Monte Carlo simulation is 
discussed in Sec.~\ref{sec43}.
Using the Monte-Carlo simulation, 
we can implement the computation of the pressure shifts for the delayed 
measurement scheme used in the Garching experiment. 
The collisional velocity spectrum and 
the number of the $2S$ atoms strongly depend on the 
delay~\cite{BeEtAl2017,YoEtAl2016,MaEtAl2013prl}.
Our approach allows us to take into account all those effects.

Finally, in addition to intra-beam collisions,
the effect of beam-background collisions is discussed
in Sec.~\ref{sec44}.
The beam-background collisions can be treated in the approximation that the 
velocity of the particles in the beam is much smaller 
than the average velocity of background gas particles.
Numerical results for intra-beam, and beam-background collisional
shifts, are given in Tables~\ref{table3} and~\ref{table4},
respectively.

For the $2S$--$4P$ experiment~\cite{BeEtAl2017},
it is shown that the possible shift in the
current configuration of the experiment is on the order of magnitude of 
$10$\,Hz, which is two orders of magnitude smaller than 
the current uncertainty of the experiment. 
In order to put this number into perspective, 
we observe that the leading uncertainties of the $2S$--$4P$ experiment~\cite{BeEtAl2017}
are the uncertainty of the Doppler shift compensation of $2.9$~kHz, 
the quantum interference shift compensation of $0.33$~kHz,
and light force shifts of $0.4$ kHz. The first-order Doppler effect also 
causes a broadening of the observed lineshape on the level of $10$~MHz. 
While the collisional effects are thus smaller than other sources 
of uncertainty in the experiment, 
they require a rather subtle analysis, as discussed here.
The model presented in this work allows us also to 
estimate the collisional shift for the 
ongoing $2S$--$6P$ experiments in the Garching laboratory.
In view of data presented in Table~\ref{table2},
the approach can easily be generalized to other transitions. 


%
%
\section*{Acknowledgments}

The authors acknowledge insightful conversations with 
Professor T W H\"{a}nsch, Th Udem and V~Debierre.
This research has been supported by the 
National Science Foundation (Grant PHY--1710856),
as well as the Missouri Research Board.
N~K~acknowledges support from DFG-RFBR grants (HA 1457/12-1 and 17-52-12016). 
A~M~acknowledges support from the 
Deutsche Forschungsgemeinschaft (DFG grant MA 7628/1-1).

\appendix

%
%
\section{First--Order van der Waals Shifts and Pressure Shift}
\label{appA}

We aim to show that the first-order van der Waals 
long-range interaction, proportional to $1/R^3$, 
does not contribute to the pressure shift 
of an atomic transition after proper
averaging over the impact parameters $\vec b$ and the 
collisional velocities $\vec v$.
To this end, we first 
recall and rewrite  Eq.~(\ref{Avg:qtoInfty}) as 
\begin{eqnarray}\label{Avg:theta}
\langle  {\rm e}^{-\mathrm{i}(\psi(t)-\psi(t-\tau))} \rangle 
= \exp\left(-\tau \int_{-\infty}^\infty a(\phi) \, [1-\ee^{-\ii \, \phi}]\, 
\mathrm{d}\phi \right)
=\ee^{-\theta(v) \tau} \,,
\end{eqnarray}
where
\begin{eqnarray}\label{theta:v}
\theta(v)= \int_{-\infty}^\infty a(\phi) \, [1-\ee^{-\ii \, \phi}]\, \mathrm{d}\phi 
=2\pi\,\mathbbm{n} \int_{-\infty}^\infty v\, b\,[1-\ee^{-\ii \, \phi}]\,
\mathrm{d} b\,.
\end{eqnarray}
In Eq.~(\ref{theta:v}),  we have  used  Eq.~(\ref{aPhi:dPhi}) to eliminate 
$a(\phi) \, \mathrm{d}\phi$. 
For any velocity distribution $P(v)$,  the quantity
\begin{eqnarray}\label{theta}
\theta= \int_0^\infty \theta(v) P(v)\, \mathrm{d} v 
\end{eqnarray}
is the impact-broadening operator (also called  the
$\theta$-operator, see Ref.~\cite{So1972}).
 It is called an ``operator'' because,
as we shall see, the phase shift $\phi$ can depend
on dipole-operator matrix elements evaluated for the 
two atoms. Being inspired by Eq.~(\ref{phi13}), 
 where a directionally averaged 
interaction, proportional to $1/R^n$, was considered,
one can go ``one step back'' and consider a 
general interaction $U(t,\vec{R})$, 
\begin{eqnarray}
\phi = \phi(v, b) =\frac{1}{\hbar}\int_{-\infty}^\infty \dd t\,  U(\vec{R}(t, \vec v, \vec b))\,,
\end{eqnarray}
where $\vec R(t) = \vec{R}(t, \vec v, \vec b ) = \vec{b} + \vec{v} \, t $ 
describes the trajectory of the atom 
with impact parameter vector~$\vec b$,
which we choose so that the closest point of 
approach is reached at $t=0$ [see Eq.~(4) of Ref.~\cite{AlGr1965}]. 
This, in particular, implies that $\vec v \cdot \vec b = 0$.
Consequently, the $\theta$-operator can be expressed as 
\begin{eqnarray}\label{thetaop}
\theta= 2\pi \mathbbm{n}\int_0^\infty v P(v) 
\int_0^\infty b\,
\left[1-\exp\left(- \frac{\mathrm{i}}{\hbar}\int_{-\infty}^{\infty} 
U(\vec{R}(t, \vec v, \vec b) \; 
\mathrm{d}t\right)\right] \,\mathrm{d} v \, \mathrm{d} b \,.
\end{eqnarray} 
The  resonance dipole-dipole
interaction  is given by 
\begin{eqnarray}\label{Eq:UtR}\fl
U(t,\vec{v},\vec{b})
&=& \frac{1}{4\pi\epsilon_0}\;\frac{\vec{d}_A(t) \cdot \vec{d}_B(t) 
-3 \,\left(\vec{d}_A(t) \cdot\widehat{R}(t) \right)
\left( \vec{d}_B(t) \cdot\widehat{R}(t)\right)}{||\vec{R}||^3} \nonumber\\
\fl& =& \frac{1}{4\pi\epsilon_0}\left[ \frac{\vec{d}_A(t) \cdot 
\vec{d}_B(t)}{\left(b^2+v^2 t^2\right)^{3/2}} 
-3 \,\frac{\left[ \vec{d}_A(t) \cdot\left(\vec{b}+\vec{v}\,t\right) \right]
\left[ \vec{d}_B(t) \cdot\left(\vec{b}+\vec{v}\, t\right) \right]}
{\left(b^2+v^2 t^2\right)^{5/2}} \right]\,.
\end{eqnarray}
where  $\vec{d}_i(t) = e\, \vec{r}_i(t)$ 
is the electric dipole operator for the atom $i=A,B$ and
$\widehat{R}=\vec{R}/||\vec{R}||$ is the unit vector along $\vec{R}(t)$. 
The pressure shift is given by the imaginary  part of the 
average value of $\theta$-operator. For resonance dipole-dipole interaction, 
we have~\cite{AlGr1965}
\begin{eqnarray}
\sigma_{\omega} =\mathrm{Im} \left<\alpha\right| \theta \left|\alpha\right>\,,
\qquad
\sigma_{\gamma} = \mathrm{Re} \left<\alpha\right| \theta \left|\alpha\right>\,.
\end{eqnarray}
Here, $\left|\alpha\right>$ is the ket corresponding to the reference 
state of the two-atom system.
Note that the operators $\vec d_i(t)$ in Eq.~(\ref{Eq:UtR}) enter the
interaction Hamiltonian in the interaction picture, i.e.,
they acquire a time dependence due to the time dependence of the 
atomic states involved in the transition.
(Strictly speaking, the exponential in Eq.~(\ref{Eq:UtR}) is
time-ordered, in the sense of an $S$-matrix element.) The question now is whether 
the first-order (in the van der Waals interaction)
effect could lead to a frequency shift.
To this end, we observe that a potential first-order 
effect is relevant only in the space of perfectly degenerate
states of the two-atom system, which can be reached 
via a dipole transition.
Within this space, however, we can replace
\begin{eqnarray}
\vec d_i(t) = \exp(\mathrm{i} H_0 \, t)  \, d_i \,  \exp(-\mathrm{i} H_0 \, t)  
\to \vec d_i \,,
\end{eqnarray}
because the operator acts in a degenerate subspace of $H_0$,
which is the unperturbed Hamilton operator of the atom.
We thus get, to first order in perturbation theory
[see Eq.~(5) of Ref.~\cite{AlGr1965}],
\begin{eqnarray}\label{Eq:IntUtR2ndO}\fl
 1-\exp\left(- \frac{\mathrm{i}}{\hbar}\int_{-\infty}^{\infty} \mathrm{d}t\,
U(t,\vec{b},\vec{v} ) \right)\nonumber\\
\fl\quad
\approx \frac{2\mathrm{i}}{4\pi\epsilon_0\,\hbar v\,b^2}
\left[ \vec{d}_A\cdot \vec{d}_B-2\left(\vec{d}_A\cdot\widehat{b}\right)
\left( \vec{d}_B\cdot\widehat{b}\right)-\left(\vec{d}_A\cdot\widehat{v}\right)
\left( \vec{d}_B\cdot\widehat{v}\right)\right] \,,
\end{eqnarray}
where $\hat b$ and $\hat v$ are the unit vectors 
in the directions of the vectors $\vec b$ and $\vec v$.
The average over angles 
of the scalar product $\left(\vec{d}_A\cdot\widehat{x}\right)
\left( \vec{d}_B\cdot\widehat{x}\right)$ is given as 
\begin{eqnarray}\label{2rhat}
\left<\left(\vec{d}_A\cdot\widehat{x}\right)
\left( \vec{d}_B\cdot\widehat{x}\right)\right>
=\frac{1}{3} \vec{d}_A\cdot \vec{d}_B\,.
\end{eqnarray}
As a result, the pressure shift in the resonance 
dipole-dipole interaction, in first-order perturbation theory, 
i.e., the average of the quantity 
\begin{eqnarray}
\vec{d}_A\cdot \vec{d}_B-2\left(\vec{d}_A\cdot\widehat{b}\right)
\left( \vec{d}_B\cdot\widehat{b}\right)-\left(\vec{d}_A\cdot\widehat{v}\right)
\left( \vec{d}_B\cdot\widehat{v}\right) \,,
\end{eqnarray}
vanishes after angular averaging over the directions of $\vec b$,
and of $\vec v$. Note that the necessity of taking this average has been implied,
but not explicitly written, in Eq.~(\ref{theta}). 
The same approach is followed in Ref.~\cite{AlGr1965}.

%
%
\section{Averaging the Cross Sections}
\label{appB}

For reference, we give some unified formulas which
illustrate the averaging procedure outlined in the discussion
surrounding Eq.~(\ref{C6:2by5:Avg}), and the
formulas for the cross sections given in Eqs.~(\ref{sigmaOmega6V}),
(\ref{xi:omega}) and~(\ref{xi:gamma}).
Recall  Eqs.~(\ref{ShiftFromXi}) and  (\ref{BroadFromXi}) for  $\omega_c$
and  $\gamma_c$ and substitute  $\xi^{(6)}_{\omega}$ and  $\xi^{(6)}_{\gamma}$
from Eqs.~(\ref{sigmaOmega6V}). 
we have 
\begin{eqnarray}
\label{PS:C6}
\omega_c&=& - 
\frac{3^{2/5} \sqrt{5-\sqrt{5}} \,
\Gamma \left(\frac{3}{5}\right) \, \Gamma \left(\frac{9}{5} \right) \,
\pi^{9/10}}{2^{7/5}} \mathbbm{n} \,
\left(\frac{k_B T}{m}\right)^{3/10} \mathrm{sgn}(C_6) 
\left(\frac{|C_6|}{\hbar}\right)^{2/5} \nonumber\\[0.1133ex]
&=&-3.79913\;  \mathbbm{n} \, \left(\frac{k_B T}{m}\right)^{3/10} 
\mathrm{sgn}(C_6)  \;\left(\frac{|C_6|}{\hbar} \right)^{2/5}\,,\\[0.1133ex]
\label{PB:C6}
\gamma_c&=&-\frac{3^{2/5} \left(\sqrt{5\,}+1\right) 
\Gamma \left(-\frac{2}{5}\right) \Gamma \left(\frac{9}{5}\right) \,
\pi^{9/10}}{2^{9/10}\times 5}
\mathbbm{n} \, \left(\frac{k_B T}{m}\right)^{3/10} 
\left(\frac{|C_6|}{\hbar} \right)^{2/5}\nonumber\\[0.1133ex]
&=&5.22906\, \mathbbm{n} \, \left(\frac{k_B T}{m}\right)^{3/10} 
\;\left(\frac{|C_6|}{\hbar}\right)^{2/5}\,.
\end{eqnarray}
It is clear from Eqs.~(\ref{PS:C6}) and (\ref{PB:C6}) that both the pressure
shift and the broadening cross-section depend on $C_6$ and $T$
according to a functional dependence of the form
$| C_6 |^{2/5} $  and $T^{3/10}$.  
The average shifts and the broadening can be written as  
\begin{eqnarray}
\omega_c = \kappa_{\omega} \,\langle |C_6|^{2/5} \rangle\,, \qquad
\gamma_c = \kappa_{\gamma} \,\langle |C_6|^{2/5} \rangle\,,
\end{eqnarray}
where
\begin{eqnarray}
\kappa_\omega =-3.79913\;  \mathbbm{n} \, \left(\frac{k_B T}{m}\right)^{3/10} 
\mathrm{sgn}(C_6)  \;\hbar^{-2/5}\,,\\[0.1133ex]
 \kappa_{\gamma} =5.22906\, \mathbbm{n} \, \left(\frac{k_B T}{m}\right)^{3/10} 
 \;\hbar^{-2/5}\,,
\end{eqnarray}
and the $\langle |C_6|^{2/5} \rangle$ average is defined in Eq.~(\ref{C6:2by5:Avg}).
The cross sections are now given as 
\begin{eqnarray}\label{Sigma:Avg}
\omega_c = \kappa_\omega \, \frac{1}{\mathbbm{M}} \,
\sum_j \mathbbm{m}_j \, \left( | C_6^{(j)} | \right)^{2/5}\,, 
\qquad
\gamma_c = \kappa_\gamma \, \frac{1}{\mathbbm{M}} \,
\sum_j \mathbbm{m}_j \, \left( | C_6^{(j)} | \right)^{2/5}\,.
\end{eqnarray}
Keeping in mind that  almost all atoms in the  background
are in  the $1S$ state, we obtain the pressure-shift 
$\omega_c $ and the pressure-broadening 
$\gamma_c$ for  
$1S$--$4P_J$ and  $1S$--$6P_J$ transitions as given in 
Eqs~(\ref{numerics_a})--(\ref{numerics_d})  above.  
One should, however,  note that 
\begin{eqnarray}
\langle | C_6 |^{2/5}  \rangle =
\frac{1}{\mathbbm{M}} \,
\sum_j \mathbbm{m}_j \, | C_6^{(j)} |^{2/5} 
\neq 
\left( \frac{1}{\mathbbm{M}} \,
\sum_j \mathbbm{m}_j \, | C_6^{(j)} | \right)^{2/5} = \langle | C_6 | \rangle^{2/5}  \,.
\end{eqnarray}
Numerically, calculations show that for the  atomic
systems under consideration here,
the difference between the two averaging procedures is 
relatively small but significant. 

%
%
\section{Deflection Radius}
\label{appC}

We assume that $R(t)$ is the time-dependent
distance between the atoms, where $b$
is the impact parameter.
Then, the distance-dependent energy shift $E$ and 
the force $F$ can be expressed as follows
[see Eqs.~(\ref{D6sign}) and~(\ref{defC6})],
\begin{eqnarray}
\fl \qquad E = -\frac{C_6}{R^6} \,,
\qquad
F = -\frac{\partial E}{\partial R} = -6 \, \frac{C_6}{R^7} \,,
\qquad
R(t) = \sqrt{ (v \,t)^2 + b^2}\,.
\end{eqnarray}
By Newton's first law, in nonrelativistic approximation,
we can write the transverse acceleration 
$a_\perp(t)$ and the transverse velocity $v_\perp(t)$ as
\begin{eqnarray}
\fl \qquad  a_\perp(t) = \frac{F \, \cos\vartheta}{m_{\rm H}} \,,
\qquad
v_\perp(t) = \int a_\perp(t) \, \dd t \,,
\qquad
\cos\vartheta = \frac{b}{ \sqrt{ (v \,t)^2 + b^2} } \,,
\end{eqnarray}
where $m_{\rm H}$ is the mass of the hydrogen atom,
approximately equal to the proton mass.
The modulus of the final transverse velocity after the collision is
\begin{eqnarray}
\fl \qquad   v_\infty = 
\left| \int_{-\infty}^\infty a_\perp(t) \, \dd t \right| = 
\frac{15 \pi}{8} \frac{C_6}{b^6 \, m_{\rm H} \, v} \,.
\end{eqnarray}
The deflection angle $\alpha$ is given by the relation
\begin{eqnarray}
\fl \qquad  \tan \alpha = \frac{v_\infty}{v} = 
\frac{15 \pi}{8} \frac{C_6}{b^6 \, m_{\rm H} \, v^2}  \,.
\end{eqnarray}
A quick calculation with $R = 100 \, a_0$ and $v = 300 \, {\rm m}/{\rm s}$
shows that the deflection angle is of the order of 
about $10^{-2}$~rad and can thus fully be neglected for 
$C_6 = 10^5 \, {\rm a.u.}$, i.e., for 
$4P$--$1S$ collisions (see Sec.~3 of Ref.~\cite{JeAdDaMaKo2018jpb1}),
but the atom is fully kicked out of its path for 
$C_6 = 10^9 \, {\rm a.u.}$,
which is the relevant range for 
$4P$--$2S$ collisions (see Sec.~4 of Ref.~\cite{JeAdDaMaKo2018jpb1}).
As explained here in Sec.~\ref{sec4}, atoms which are kicked 
out from the beam only contribute to the experimental background. 
Yet, as Table~\ref{table3} shows, the contribution of 
$4P$--$2S$ collisions to the collisional frequency shift is 
much smaller than that of $4P$--$1S$,
so that an over-estimation of the former has negligible 
effect on the total estimate of the collisional frequency shift.

A last remark is in order.
For reference, we can point out that by setting $\tan \alpha = 1$, 
we can define a ``deflection radius'' $\rho_{\rm D}$, 
\begin{eqnarray}
\fl \qquad  \rho_{\rm D} = 
\left( \frac{15 \pi}{8} \frac{C_6}{m_{\rm H} \, v^2} \right)^{1/6} \,,
\end{eqnarray}
which is the radius below which the deflection of an 
incoming atom becomes significant; it is the analogue 
of the well-known Weisskopf radius which describes the 
onset of a significant phase shift
during a collision.
For our experimental conditions, a numerical estimate shows
that the deflection radius and the Weisskopf radius are of the 
same order-of-magnitude, implying some of the 
atoms otherwise affected by collisional are being kicked out of the beam.
As explained in Sec.~\ref{sec4}, the estimates of the 
collisional frequency shifts obtained here, thus constitute upper limits
for the effect in the Garching experiment~\cite{BeEtAl2017}.

\hspace*{1cm}

\end{document}